\documentclass[prd,twocolumn,superscriptaddress,nofootinbib]{revtex4}
\usepackage{bm,amsmath,amssymb,epsfig,natbib}
\usepackage[hypertex]{hyperref}
\usepackage{ifthen}

\begin{document}

\newcommand{\lexact}{{l_\text{exact}}}
\newcommand{\llow}{{l_\text{low}}}

\newcommand{\Li}{\text{Li}}

\newcommand {\mDh} { \hat{\bm{D} } }
\newcommand {\mUh} { \hat{\bm{U} } }

\newcommand{\nside}{N_{\text{side}}}
\newcommand{\arcmin}{\text{arcmin}}
\newcommand{\begm}{\begin{pmatrix}}
\newcommand{\enm}{\end{pmatrix}}
\newcommand{\threej}[6]{{\begm #1 & #2 & #3 \\ #4 & #5 & #6 \enm}}
\newcommand{\Coff}{{\hat{C}_l^\off}}
\newcommand{\lmax}{l_\text{max}}
\newcommand{\lmin}{l_{\text{min}}}
\newcommand{\fsky}{f_{\text{sky}}}
\newcommand{\off}{{\text{off}}}
\newcommand{\chieff}{\chi^2_{\text{eff}}}
\renewcommand{\ell}{l}
\renewcommand{\bar}[1]{#1}
\newcommand{\nHI}{{n_{HI}}}
\newcommand{\clh}{\mathcal{H}}
\newcommand{\ud}{{\rm d}}

\def\eprinttmp@#1arXiv:#2 [#3]#4@{
\ifthenelse{\equal{#3}{x}}{\href{http://arxiv.org/abs/#1}{#1}}{\href{http://arxiv.org/abs/#2}{arXiv:#2} [#3]}}

\renewcommand{\eprint}[1]{\eprinttmp@#1arXiv: [x]@}
\newcommand{\adsurl}[1]{\href{#1}{ADS}}
\renewcommand{\bibinfo}[2]{\ifthenelse{\equal{#1}{isbn}}{%
\href{http://cosmologist.info/ISBN/#2}{#2}}{#2}}

\newcommand\ba{\begin{eqnarray}}
\newcommand\ea{\end{eqnarray}}
\newcommand\be{\begin{equation}}
\newcommand\ee{\end{equation}}
\newcommand\lagrange{{\cal L}}
\newcommand\cll{{\cal L}}
\newcommand\clx{{\cal X}}
\newcommand\clz{{\cal Z}}
\newcommand\clv{{\cal V}}
\newcommand\clo{{\cal O}}
\newcommand\cla{{\cal A}}
\newcommand{\uD}{{\mathrm{D}}}
\newcommand{\calE}{{\cal E}}
\newcommand{\calB}{{\cal B}}
\newcommand{\curl}{\,\mbox{curl}\,}
\newcommand\del{\nabla}
\newcommand\Tr{{\rm Tr}}
\newcommand\half{{\frac{1}{2}}}
\renewcommand\H{{\cal H}}
\newcommand\K{{\rm K}}
\newcommand\mK{{\rm mK}}
\newcommand{\clk}{{\cal K}}
\newcommand{\bq}{\bar{q}}
\newcommand{\bv}{\bar{v}}
\renewcommand\P{{\cal P}}
\newcommand{\numfrac}[2]{{\textstyle \frac{#1}{#2}}}
\newcommand{\la}{\langle}
\newcommand{\ra}{\rangle}
\newcommand{\rar}{\rightarrow}
\newcommand{\Rar}{\Rightarrow}
\newcommand\gsim{ \lower .75ex \hbox{$\sim$} \llap{\raise .27ex \hbox{$>$}} }
\newcommand\lsim{ \lower .75ex \hbox{$\sim$} \llap{\raise .27ex \hbox{$<$}} }
\newcommand\bigdot[1] {\stackrel{\mbox{{\huge .}}}{#1}}
\newcommand\bigddot[1] {\stackrel{\mbox{{\huge ..}}}{#1}}
\newcommand{\Mpc}{\text{Mpc}}
\newcommand{\Al}{{A_l}}
\newcommand{\Bl}{{B_l}}
\newcommand{\eAl}{e^\Al}
\newcommand{\ix}{{(i)}}
\newcommand{\ixp}{{(i+1)}}
\renewcommand{\k}{\beta}
\newcommand{\HD}{\mathrm{D}}
\newcommand{\mCh}{\hat{\bm{C}}}
\newcommand{\mCf}{{{\bm{C}}_{f}}}
\newcommand{\mCXf}{{{\bm{C}}_{Xf}}}
\newcommand{\mMXf}{{{\bm{M}}_{f}}}
\newcommand{\Cfl}{{C_f}_l}
\renewcommand{\vec}{\text{vec}}
\newcommand{\muK}{\mu \rm{K}}

\newcommand{\var}{\text{var}}
\newcommand{\cov}{\text{cov}}
\newcommand{\bias}{\text{bias}}

\newcommand{\vecp}{\text{vecp}}
\newcommand{\vecL}{\text{vecl}}

\newcommand{\mCfl}{{\mC_{f}}_l}
\newcommand{\mCgl}{{\mC_{g}}_l}

\newcommand{\Ch}{\hat{C}}
\newcommand{\Bt}{\tilde{B}}
\newcommand{\Et}{\tilde{E}}
\newcommand{\bld}[1]{\mathrm{#1}}
\newcommand{\mLambda}{\bm{\Lambda}}
\newcommand{\mA}{\bm{A}}
\newcommand{\mB}{\bm{B}}
\newcommand{\mBp}{\mB_n}

\newcommand{\mC}{\bm{C}}
\newcommand{\mD}{\bm{D}}
\newcommand{\mE}{\bm{E}}
\newcommand{\mF}{\bm{F}}
\newcommand{\mg}{\bm{g}}

\newcommand{\mQ}{\bm{Q}}
\newcommand{\mU}{\bm{U}}
\newcommand{\mX}{\bm{X}}
\newcommand{\mV}{\bm{V}}
\newcommand{\mP}{\bm{P}}
\newcommand{\mR}{\bm{R}}
\newcommand{\mW}{\bm{W}}
\newcommand{\mI}{\bm{I}}
\newcommand{\mH}{\bm{H}}
\newcommand{\mM}{\bm{M}}
\newcommand{\mN}{\bm{N}}
\newcommand{\mMh}{\hat{\mM}}

\newcommand{\mY}{\bm{Y}}

\newcommand{\vs}{\mathbf{s}}
\newcommand{\vshat}{\hat{\mathbf{s}}}

\newcommand{\vv}{\mathbf{v}}
\newcommand{\vd}{\mathbf{d}}
\newcommand{\vC}{\mathbf{C}}
\newcommand{\vT}{\mathbf{T}}

\newcommand{\mS}{\bm{S}}
\newcommand{\mzero}{\bm{0}}
\newcommand{\mL}{\bm{L}}
\newcommand{\btheta}{\bm{\theta}}
\newcommand{\bphi}{\bm{\psi}}
\newcommand{\va}{\mathbf{a}}
\newcommand{\vX}{\mathbf{X}}
\newcommand{\vchi}{\bm{\chi}}

\newcommand{\vXh}{\hat{\vX}}

\newcommand{\vS}{\mathbf{S}}
\newcommand{\vm}{\mathbf{m}}
\newcommand{\vn}{\mathbf{n}}

\newcommand{\vN}{\mathbf{N}}
\newcommand{\vXhat}{\hat{\mathbf{X}}}
\newcommand{\vb}{\mathbf{b}}
\newcommand{\vA}{\mathbf{A}}
\newcommand{\vAt}{\tilde{\mathbf{A}}}
\newcommand{\ve}{\mathbf{e}}
\newcommand{\vE}{\mathbf{E}}
\newcommand{\vB}{\mathbf{B}}
\newcommand{\vl}{\mathbf{l}}
\newcommand{\vp}{\mathbf{p}}
\newcommand{\vP}{\mathbf{P}}

\newcommand{\vXf}{\mathbf{X}_f}
\newcommand{\vEt}{\tilde{\mathbf{E}}}
\newcommand{\vBt}{\tilde{\mathbf{B}}}
\newcommand{\vEw}{\mathbf{E}_W}
\newcommand{\vBw}{\mathbf{B}_W}
\newcommand{\vx}{\mathbf{x}}
\newcommand{\vXt}{\tilde{\vX}}
\newcommand{\vXb}{\bar{\vX}}
\newcommand{\vTb}{\bar{\vT}}
\newcommand{\vTt}{\tilde{\vT}}
\newcommand{\vY}{\mathbf{Y}}
\newcommand{\vBwr}{{\vBw^{(R)}}}
\newcommand{\RW}{{W^{(R)}}}
\newcommand{\mUt}{\tilde{\mU}}
\newcommand{\mVt}{\tilde{\mV}}
\newcommand{\mDt}{\tilde{\mD}}

\newcommand{\healpix}{HEALPix}

\title{Properties and use of CMB power spectrum likelihoods}

\author{Samira Hamimeche}
\email{samira@ast.cam.ac.uk}
 \affiliation{Institute of Astronomy and Kavli Institute for Cosmology, Madingley Road, Cambridge, CB3 0HA, UK.}

\author{Antony Lewis}
\homepage{http://cosmologist.info}
\affiliation{Institute of Astronomy and Kavli Institute for Cosmology, Madingley Road, Cambridge, CB3 0HA, UK.}

\date{\today}

\begin{abstract}
Fast robust methods for calculating likelihoods from CMB observations on small scales generally rely on approximations based on a set of power spectrum estimators and their covariances. We investigate the optimality of these approximation, how accurate the covariance needs to be, and how to estimate the covariance from simulations. For a simple case with azimuthal symmetry we compare optimality of hybrid pseudo-$C_l$ CMB power spectrum estimators with the exact result, indicating that the loss of information is not negligible, but neither is it enough to have a large effect on standard parameter constraints. We then discuss the number of samples required to estimate the covariance from simulations, with and without a good analytic approximation, and assess the use of shrinkage estimators. Finally we discuss how to combine an approximate high-$l$ likelihood with a more exact low-$l$ harmonic-space likelihood as a practical method for accurate likelihood calculation on all scales.
\end{abstract}

\maketitle

\pagenumbering{arabic}

\section{Introduction}

The Cosmic Microwave Background (CMB) appears to be isotropic and Gaussian to a good approximation, and hence allows robust statistical constraints to be placed on a variety of cosmological parameters. However high resolution observations such as those from the Planck satellite produce sky maps with many millions of pixels, for which performing an exact likelihood analysis becomes numerically very difficult. Most data analyses therefore use fast robust approximations based on power spectrum estimators~\cite{Bond:1998qg,Verde:2003ey}. While these are expected to be suboptimal at some level, they should be unbiased on average. The main advantage of a fast method is that multiple simulations can be performed to assess in detail the propagation of errors from numerous instrumental processes into the final results, and hence in practice may also be significantly more reliable than an in-principle more optimal method.

In a previous paper we investigated in detail various approximations for calculating cosmological parameter likelihoods from high-resolution CMB power spectrum estimators, and showed that indeed good approximations can be found that produce unbiased results~\cite{Hamimeche:2008ai}. These approximations effectively transform a set of power spectrum estimators so the likelihood of a theoretical power spectrum can be written in a Gaussian form, then evaluate this Gaussian function using an estimate of the power spectrum estimator covariance. In simple cases this covariance can be calculated to reasonable accuracy using analytic approximations, though ideally it should be assessed by performing large numbers of full data-analysis simulations, fully accounting for noise, sky cuts, noise correlations, beam effect, map-making errors, etc. Additional complications may additionally be accounted for by modifying the theory power spectrum, for example due to beam or foreground uncertainty modes.

In this paper we aim to assess how much information is being lost by using a fast pseudo-$C_l$ method compared to a more optimal method (e.g. maximum likelihood or Gibbs sampling~\cite{Wandelt:2003uk}), and whether the increase in error bars from using a suboptimal method has a significant effect on cosmological parameters. We then assess how to estimate the covariance from simulations, and how many simulations are required under various assumptions. We show how information from approximate models can be combined with simulations by using shrinkage estimators or by fitting model parameters. Finally we suggest how an approximate power spectrum likelihood at high $l$ can be combined with a more exact low-$l$ likelihood for evaluating the total likelihood, consistently accounting for the contribution of high $l$ power to the low-$l$ mode variance.

 We focus on nearly full-sky observations, for example from the WMAP or Planck satellites, where only a small fraction ($\sim 15\%$) of the sky is cut out due to foreground or point-source contamination. For simplicity we discuss mainly the CMB temperature, which is what drives the parameter constraints, though in the appendix we give a general harmonic method for calculating the low-$l$ likelihood in harmonic space. Since the paper is only likely to be of interest to experts in the field, we refer to the extensive literature for introductory material.

\section{How optimal are pseudo-$C_l$ likelihoods?}

We focus on hybrid pseudo-$C_l$ power spectrum estimators, constructed by combining sets of
pseudo-$C_l$ estimators from maps with different weightings. For details of how the pseudo-$C_l$ estimators are constructed, and how the covariance can be estimated, see Refs.~\cite{Tegmark:1996qt,Wandelt:2000av,Hivon:2001jp,Hansen:2002zq,Efstathiou:2003dj,Brown:2004jn,Hinshaw:2006ia,Efstathiou:2006eb,Smith:2006vq,Hamimeche:2008ai}.
When the sky is noise dominated, the pseudo-$C_l$ estimator with inverse-noise weighting is optimal;
when the noise is negligible a uniform weighting is optimal.
Combining different pseudo-$C_l$ estimators with different weightings gives hybrid estimators that interpolate smoothly between the regimes, giving significantly smaller error bars at all $l$ than a single weight function could~\cite{Efstathiou:2003dj}.

Over the years various code comparisons and tests have indicated that pseudo-$C_l$ estimator are in fact rather good at high $l$, in that their variance is typically within
$\clo(10\%)$ of that expected for a maximum likelihood estimator (e.g. Refs.~\cite{Hivon:2001jp,Efstathiou:2006eb}). However the degree of sub-optimality will depend on the noise and sky cut under consideration. Due to its scanning strategy, the Planck satellite will have highly anisotropic noise, and we would like to assess whether this is likely to be a problem for pseudo-$C_l$ methods. Here we perform a comparison in the case of a strongly anisotropic noise, but with azimuthal symmetry so that the optimal Fisher errors can be calculated exactly for comparison. Though this situation is clearly unrealistic, it should give a good idea of the amount of suboptimality that can be expected in practice. Our main concern is the anisotropy of the noise, so we shall consider the full sky; the main advantage of doing this is that the covariance of the pseudo-$C_l$ estimators can then be calculated accurately analytically, so that our comparison is assessing the information loss due to sub-optimal handling of noise anisotropy, rather than depending on errors in the covariance calculation. We discuss the effects of covariance errors in a later section. We focus on the high-$l$ temperature power spectrum since this dominates the parameter constraints from the Planck satellite; for discussion of how to improve polarization pseudo-$C_l$ estimators see Refs.~\cite{Smith:2005ue,Smith:2006vq}.

\subsection{Comparing the Fisher matrix and the pseudo-$C_{l}$'s variance in the full sky}

The Cram\'er-Rao inequality states that the inverse of the Fisher information is a lower bound on the variance of any unbiased estimator, so we use the Fisher matrix to quantify the errors on the power spectrum that one could hope to get using an optimal method. For a Gaussian likelihood function $\cll$, the  Fisher matrix, $\mF$, for the power spectrum $C_l$ at some fiducial model is
given by (see e.g. Ref.~\cite{Oh:1998sr})
\begin{align}
\begin{split}
F_{ll'} & = -\left \langle \left( \frac{\partial^2}{\partial C_l \partial C_{l'}} \right) \ln \mathcal{L} \right \rangle \\
& = \frac{1}{2} \text{Tr} \left[ \mC^{-1} \frac{\partial \mC}{\partial C_l} \mC^{-1} \frac{\partial \mC}{\partial C_{l'}} \right],
\end{split}
\end{align}
where $\mC=\mS+\mN$, with $\mS$ and $\mN$ being the signal and noise covariances, respectively.
For a statistically isotropic signal on the full sky $\mS$ is diagonal in harmonic space, and for pixel-uncorrelated noise $\mN$ can be calculated easily in terms of the pixel noise variance. We make the assumption of azimuthal symmetry so that $N_{l'm'l m} = \delta_{mm'} N_{l'ml m}$, which makes the Fisher matrix numerically tractable by putting it into block diagonal form (with blocks for each $m$).
The Fisher matrix then reads:
\begin{align}\begin{split}
& F_{ll'} = \frac{1}{2} \sum_m \left( [\mC^{-1}]_{ll'}^{(m)} \right)^{2}, \label{FisherMatrix}
\end{split}\end{align}
which is simply half the square of the inverse of the covariance matrix (signal plus noise) summed over $m$.

We compare this with the errors expected using pseudo-$C_l$ power spectrum estimators.
For weight functions labelled by $i$ and $j$ there are a set of pseudo-$C_l$ estimators $\hat{C}_l^{ij}$,
and a covariance
\begin{equation} \label{CovwithNoise}
M_{ll'}^{ijkl} \equiv \langle \Delta \Ch_{l}^{ij} \Delta \Ch_{l'}^{kl} \rangle.
\end{equation}
Explicit expressions for the covariances are given in e.g. Ref.~\cite{Hamimeche:2008ai}.
The hybrid pseudo-$C_{l}$ estimators of Ref.~\cite{Efstathiou:2003dj,Hamimeche:2008ai} are a combination of pseudo-$C_{l}$'s with different combinations of weights.  In the low noise regime, the best weight function, $\omega(\Omega)$ is close to uniform in order to minimize the cosmic variance.  However, in the high noise regime, the weight function is proportional to the inverse-noise to reduce noise itself~\cite{Efstathiou:2003dj}.  Therefore, a natural choice would be to choose a combination of results from uniform and inverse-noise weight functions. Here we consider the simplest case of combining the two estimators calculated
 separately from the two weighted maps.
The full hybrid covariance matrix then reads:
\begin{equation}\label{CovHybrid}
M^h_{ll'} \equiv \langle \Delta \Ch_{l}^{h} \Delta \Ch_{l'}^{h} \rangle = \left( \sum_{AB} H_{ll'}^{AB}\right)^{-1} .
\end{equation}
where $H_{ll'}^{AB}$ is the inverse of the covariance matrix
 $\langle \Delta \Ch_{l}^{A} \Delta \Ch_{l'}^{B} \rangle$, with $A$ and $B$ denoting the different weight functions used for the $\hat{C}_l$ contributing to the hybrid estimator.

\begin{figure}[htbp]  \centering
\epsfig{file=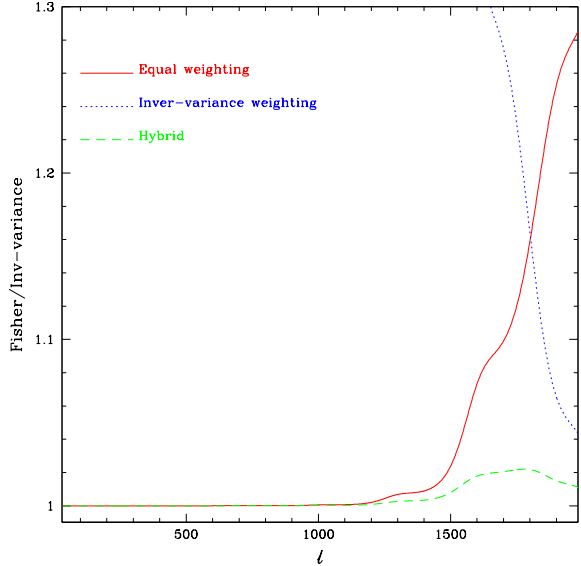, width=9cm}
\caption{A comparison between the Fisher errors and the inverse of the hybrid estimator variance for the different weighting combinations considered.  The red (solid), blue (dotted) and green (dashed) curves represent the ratios of Fisher errors to the inverse variance of uniform noise weighting, inverse-variance noise weighting and hybrid, respectively.}
\label{fig:TFisherHybrid}
\end{figure}
We can now proceed to compare the errors on individual $C_l$ and their correlations. However for parameter estimation we are not just interested in individual $l$: the $C_l$ are smooth functions of $l$, and the effects of parameters enter in a smooth way over a range of $l$. We are therefore more interested in the constraint on the amplitude over some smoothing scale, as this will more closely determine how well we can constrain different parameters. The constraint over a range of scales depends on the correlations of different individual $l$, so considering a range of $l$ also has the advantage of checking any effects due to correlations.

We therefore also evaluate the error of an amplitude parameter, $A$, over some range in $\Delta l$, so that the power spectrum over the range $\Delta l$ is given by $AC_l^{\rm{in}}$, and elsewhere by $C_l^{\rm{in}}$, for some fiducial model $C_l^{\rm{in}}$,.
The Fisher matrix for amplitude $A$ over $\Delta l$ is then
\begin{align}\begin{split}
\mF_{AA} &=  \frac{1}{2} \sum_{m} \sum_{l=|m|}^{l_{max}}
\sum_{l'=|m|}^{l_{max}} [\mC^{-1}]_{ll'}^{2} C_{l}^{\rm{in}} C_{l'}^{\rm{in}} \\
&\times
\left. \left\{ \begin{array}{cc} 1 & \text{if $l$ and $l'$ in $(\Delta l)$} \\ 0 & \text{otherwise} \end{array} \right. \right. .
\end{split}\end{align}

This can be compared to using the pseudo-$C_l$ estimators in a Gaussian likelihood approximation:
\begin{equation}\label{gaussfisher1}
-2 \ln \mathcal{L}_{f}(C_l) =  \sum_{ll'} (\Ch_{l} - C_{l}) [\mM_{f}^{-1}]_{ll'} (\Ch_{l'} - C_{l'}),
\end{equation}
where $[\mM_{f}]$ is the covariance for a fiducial model. Differentiating twice w.r.t $A$ gives:
\begin{equation}\label{gaussfisher2}
-\frac{\partial^2}{\partial A ^2}  (\ln \mathcal{L}_{f})  = \sum_{ll'} C_{l}^{in} [\mM_{f}^{-1}]_{ll'} C_{l'}^{in},
\end{equation}
where the sum is over the $l$ in the bin being considered. Note that we are considering the case of generating estimators at each $l$, and then using the covariance to estimate the power over a range of $l$. This is not the same as making a binned $C_l$ estimator (depending on some window function range of $l$), which may well have different properties. Full sky observations such as WMAP and Planck are normally analysed into individual $C_l$ estimators, the case we consider, though there may be advantages to also considering binned estimators.

To make an error comparison, we use an azimuthally averaged (in ecliptic coordinates) version of the noise expected
for the Planck satellite~\cite{Ashdown:2007ta}  and consider an isotropic 7arcmin-fwhm Gaussian beam.

We compare the inverse of the hybrid variance (the diagonal elements of Eq.~\eqref{CovHybrid}) to the Fisher errors (the diagonals of Eq.~\eqref{FisherMatrix}) in Fig.~\ref{fig:TFisherHybrid}. The figure also shows the uniform and inverse-weighted estimators separately.
Although the variance of the uniform noise weighting is almost identical to the fisher error at low $l$, it deviates considerably when noise dominates.  On the other hand, the variance of the inverse-variance noise weighting case remains poor even at high $l$ where noise is starting to dominate over the signal.  The hybrid estimators significantly improves the errors, giving results within a few percent of the ideal Fisher errors at all $l$.

\begin{figure}[t]  \centering
\epsfig{file=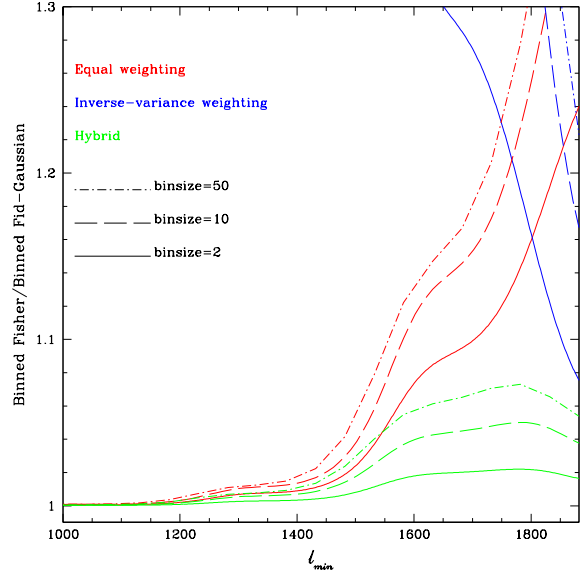, width=9cm}
\caption{A comparison of the errors in the power over a range $\Delta l=2$ (solid line), $\Delta l=10$ (dashed line) and $\Delta l=50$ (dot-dashed line), for bands starting at $\lmin$. The red, blue and green curves represent the ratios of Fisher errors to the inverse variance of uniform noise weighting, inverse-variance noise weighting and hybrid, respectively.  }
\label{fig:TFisherHybridBinning}  \end{figure}


Fig.~\ref{fig:TFisherHybridBinning} compares the results for the amplitude over a range $\Delta l=2, 10, 50$. In this case a larger range of $l$ gives relatively worse hybrid errors compared to the Fisher errors, though the hybrid errors are still within $10\%$ at all $l$. One interpretation of the effect of bin size might be that the optimal result is combining correlation information between different scales significantly more efficiently than the hybrid estimator. This is perhaps not surprising since the Fisher result `knows' about correlations between all the $l$ and $m$ modes individually, whereas the hybrid approximation has compressed the correlation information into a covariance matrix accounting only for $l$ correlations. The hybrid estimator could be further improved by including more pseudo-$C_l$ estimators in the mix if greater optimality is desired, for example the uniform-inverse weight estimator; however as we see below the level of optimality found here is likely to be sufficient in most cases.


\subsection{Effect on parameter estimation}
\label{HybridEffect}

\begin{figure}[t]  \centering
\epsfig{file=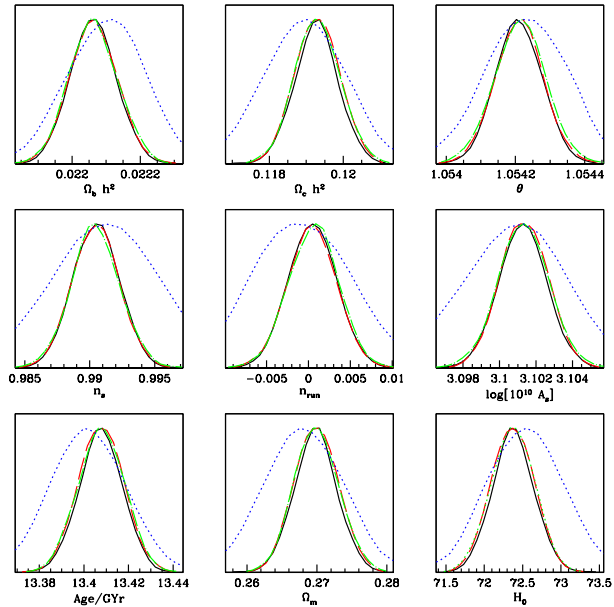, width=9cm}
\caption{Parameter constraints from a single idealized Plank-like simulation with anisotropic noise.  The 1-dimensional marginalized posteriors are from the fiducial Gaussian approximation with the covariance being from the Fisher matrix or the pseudo-$C_l$ estimator covariance.  The black (solid) line uses the Fisher covariance, blue (dotted) is the inverse-noise weighted estimator, green (dot-dashed) is with uniform weight, and red (dashed) uses the hybrid estimator.
The results are very consistent as constraints are driven by scales where the uniform weighting is nearly optimal. }
\label{fig:ParamsCompare}  \end{figure}

Since the hybrid pseudo-$C_l$ may be suboptimal at up to the $10\%$ level, it is also useful to directly check the effect on parameter estimation.  We use full-sky Planck-like simulations with azimuthal symmetry as previously described.  We consider a standard 6-parameter $\Lambda$CDM model (with running of the spectral index, but the optical depth fixed since we exclude $l\le 30 $ where a more optimal likelihood analysis is possible).  For simplicity and for a quick check, we consider the fiducial Gaussian likelihood approximation since it depends on a pre-computed covariance matrix.  We calculate the likelihood with the Fisher covariance (Eq.~\eqref{FisherMatrix}) and with the hybrid covariance (Eq.~\eqref{CovHybrid}) using the CosmoMC parameter estimation code to sample from the posterior parameter distribution~\cite{Lewis:2002ah}. We use a fiducial analytic model for the $\hat{C}_l$, corresponding to averaging the log-likelihood over realizations.

Fig.~\ref{fig:ParamsCompare} shows the results, which are very similar. This is not a surprise since Planck can measure many acoustic peaks in the CMB power spectrum, so most parameters are well constrained from the cosmic-variance limited region where the hybrid estimator is close to optimal.
 The effect could be somewhat larger in extended parameter spaces where models differ significantly only over the region where the hybrid estimator is significantly sub-optimal.

\section{How to calculate the covariance?}

In the simple cases we have considered above the estimator covariance can be calculated accurately analytically. In more realistic cases this is unlikely to be the case: even a simple sky cut renders the commonly-used covariance approximates accurate at only the $\clo(10\%)$ level, depending on apodization (see e.g.~\cite{Lewis:2008wr}). Ideally we would like to be able to evaluate the covariance (in some fiducial model) directly from simulations of the full time stream, including all relevant instrumental effects. This is the approach adopted by the MASTER/xFASTER pipelines used by many CMB observations with data over only a small part of the sky~\cite{Hivon:2001jp,Reichardt:2008ay,Jones:2005yb}; the advantage of a Monte-Carlo approach is that numerous complicated effects can be included straightforwardly, where an analytic result would be intractable. It should however be remembered that if the likelihood approximation is derived to be good for near-isotropic-Gaussian fields, including systematics in the covariance may invalidate the likelihood approximation. For example beam uncertainty modes effectively scale the theory power spectrum coherently over a large range of scales, so beam uncertainties may be better accounted for by including them as extra parameters affecting the theory $C_l$ in the parameter estimation code, rather than as part of the likelihood function for beam-uncertain estimators~\cite{Lewis:2008wr}.

The maximum likelihood estimator of the covariance of a set of zero-mean samples of a Gaussian vector $\vn$ is
\begin{equation}
\mMh = \frac{1}{n} \sum_i \vn_i \vn_i^T,
\end{equation}
where the sum is over the $n$ independent samples. If our only knowledge about the covariance comes from simulations, we should account for the sampling uncertainty in the true covariance by marginalizing over the probability distribution of the true covariance given the Monte-Carlo estimate; see appendix~\ref{appcov} where we give mathematical details in an idealized case.

If there are a large number of samples, the uncertainty in the true covariance may be negligible, so the Monte-Carlo covariance can be used directly, but the question is: how many samples do you need? There is a basic lower limit of $n\ge p$, where $p$ is the dimensionality of the covariance; this is required to have $p$ independent modes sampled, and hence for the covariance estimate to be invertible. Already in the case of the power spectrum estimators this is a non-trivial requirement: for WMAP temperature $p\sim 10^3$, but due to the very high numerical cost they were only able to do a dozen or so full time-stream simulations. For the distribution of the true covariance given the estimator to be normalizable we actually need $n > 2p$, and for fractional accuracy better than $\alpha$ on the error bars, we require $n\agt 2p/\alpha$. Indeed to have the numerical value of the $\chi^2$ accurate, one needs $n \gg p^2$, which is getting to be very time consuming for un-binned $C_l$ estimators, even for simplified map-level simulations. However in practice it is not usually required to have the $\chi^2$ accurate, as long as the variation under changes in parameters is accurate, and hence the correct parameter constraints are obtained.  For further discussion see Appendix~\ref{appcov} and Refs.~\cite{Gupta99,Hartlap:2006kj}.

In a realistic situation we may have some idea of the covariance from approximate analytical results, but would also like to calibrate it from simulations to account for complications that are not included in the analytical model. An estimator can use both the approximation and the simulations. Perhaps the simplest case to consider is where there is a prior estimate that one considers to be about as accurate as that from $q$ simulations. Then adding the information from $n$ actual simulations, the best estimate of the true covariance will be the weighted sum of the two covariance estimates (see appendix~\ref{appcov}). This is a simple example of a shrinkage estimator: it `shrinks' noisy simulated estimators towards some prior target, giving a new estimator that should be better than both individually. In the limit of many simulations the new estimator is simulation dominated; with few simulations it is prior dominated.

\subsection{Shrinkage estimators}

Shrinkage estimators were originally introduced for the situation in which there are many fewer samples than numbers of dimensions, so that the maximum-likelihood covariance estimate is not even invertible. This is a common situation in many problems, and shrinkage provides a method to regularize the estimate in a well-defined way so that it is invertible. Here we would ideally also like to have very few simulations, but we also require good accuracy of the answer, and may prefer to generate more samples than have an inaccurate answer. In this section we investigate whether shrinkage estimators are useful for CMB likelihoods.

A shrinkage estimator $ \bf{s}^*$ for a quantity $\vs$ is constructed from a linear combination of
some `target' $\bf{t}$ and an estimator
$\vshat$ calculated from samples~\cite{Ledoit03,Schafer05,Pope:2007vz}:
\begin{equation}
\bf{s}^*=\lambda \bf{t} + (1-\lambda)\vshat,
\end{equation}
with $\lambda$, the shrinkage intensity, being in the range 0 to 1.  The target can either be a fixed prior, or
it could some approximate estimate from the data, or some combination.
If $\lambda=0$ then the shrinkage estimate equals the unrestricted estimate, $\vshat$.  If, however, $\lambda=1$ then the target estimate, $\bf{t}$, is recovered.  Therefore, the main advantage of this weighted combination is to form a regularized estimator that outperforms both estimators individually.  The shrinkage intensity is chosen in a way that
would optimize the estimator, for example by minimizing the mean-squared error:
\begin{equation}
\text{MSE}(\lambda)=\la\sum_l(s^*_l - s_l)^2\ra,
\label{MSE}
\end{equation}
with the angle brackets being the expectation value.  Ledoit and Wolf~\cite{Ledoit03} derived a analytic solution
for $\lambda$ to minimize this error:
\begin{equation}\label{ShrinkageIntensity}
\lambda^* = \frac{\sum_l[ \var(\hat{s}_l) - \cov(t_l,\hat{s}_l) - \bias(\hat{s}_l)\la t_l-\hat{s}_l\ra]}
{\sum_l \la(t_l-\hat{s}_l)^2\ra}.
\end{equation}
For a practical use of Eq.~\eqref{ShrinkageIntensity}, Ref.~\cite{Schafer05} suggest replacing
the covariance, variance and bias with their unbiased samples estimates $(\hat{\cov}, \hat{\var}, \hat{\bias})$.
 If $\hat{s}$ is an unbiased estimator, then the $\bias$ term may be omitted. The
  new expression of the shrinkage intensity then reads:
\begin{equation}
\hat{\lambda}^* = \frac{\sum_l [\hat{\var}(\hat{s}_l) - \hat{\cov}(t_l,\hat{s}_l) ]}{\sum_l (t_l-\hat{s}_l)^2}.
\end{equation}
In practice we use $\text{min}(1,\hat{\lambda}^*)$.

The shrinkage method can be applied directly to estimate the covariance matrix when $\vs$ is taken to be a vector of the distinct elements of the covariance matrix.
If we have $n$ measurements of a data vector
 $\bf{x}$ of size $p$, and we assume that $x_l^{(k)}$ is the $k^{th}$ measurement of the $l^{th}$ element of $\bf{x}$, then the empirical mean of the variable $x_l$ is:
\begin{equation}
\overline{x}_l = \frac{1}{n} \sum_{k=1}^{n} x_{l}^{(k)}.
\end{equation}
The unbiased empirical covariance, $\hat{\bf{S}}$ is given by:
\begin{equation}
\hat{S}_{ll'} = \hat{\cov}(x_l,x_{l'}) = \frac{n}{n-1} \overline{W}_{ll'},
\end{equation}
where we have set:
\begin{eqnarray}
\overline{W}_{ll'} = \frac{1}{n} \sum_{k=1}^n W_{ll'}^{(k)},\\
W_{ll'}^{(k)} = (x_{l}^{(k)} - \overline{x}_{l})(x_{l'}^{(k)} - \overline{x}_{l'}).
\end{eqnarray}
Similarly, the required variance can be estimated as
\begin{equation}
\hat{\var}(S_{ll'}) = \frac{n}{(n-1)^3}\sum_{k=1}^{n}(W_{ll'}^{(k)} - \overline{W}_{ll'})^2.
\end{equation}
If the target is fixed (does not depend on the data) then the covariance term vanishes; if not, it should be included to account for the correlation. We estimate the shrinkage intensity by replacing $\vshat$ and $\bf{t}$ in Eq.~\eqref{ShrinkageIntensity}
with $\hat{\bf{S}}$ and $\bf{T}$, where $\bf{T}$ is some arbitrary target matrix.  Therefore, the optimized
 shrinkage covariance now takes the form:
\begin{equation}
\bf{M} = \hat{\lambda}^* \bf{T} + (1 - \hat{\lambda}^*)\hat{\bf{S}}.
\end{equation}
The shrinkage estimator is designed to use information from the sample estimator roughly in proportion to how relatively accurate it is. If we have a target that is say $10\%$ accurate, $\hat{\bf{S}}$ needs to have $\alt 10\%$ accuracy to significantly improve on the target. For high precision results, the estimator needs to be accurate, so the shrinkage estimator requires the same order of number of samples as the direct accurate estimator. It simply has the advantage of being slightly superior, and not breaking down rapidly as the number of samples is decreased.

Note that it is not at all clear in the current context that minimizing the Frobenius norm of the error (Eq.~\ref{MSE}) is the best thing to do. Indeed if the diagonal and non-diagonal elements of the covariance have very different distributions, it is not a good idea to use a shrinkage estimator derived by treating them on equal terms. One can however equally well apply a shrinkage estimator to subsets of parameters separately. Since the off-diagonal terms are generally small, the number of simulations required to estimate them accurately is very large: for a given number of samples the shrinkage intensity $\lambda^*$ will be significantly larger for the off-diagonal than the diagonal components. Also note that the shrinkage intensity $\lambda^*$ is not independent of the scaling of the elements (e.g. using $C_l$ or $l(l+1)C_l$ gives different results). We therefore apply the estimator to the covariance matrix after taking out the approximate $C_l C_{l'}$ scaling of the elements (e.g. normalizing so that the target has unit diagonal).

\begin{figure}
\begin{center}
\epsfig{figure=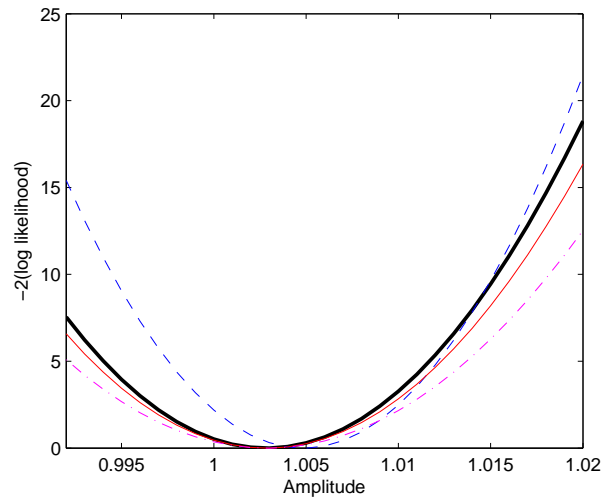,width=8cm}
\caption{Example log likelihood for the power spectrum amplitude over $100 < l\le 700$ in a typical realization when using the exact covariance (thick solid), a prior guess diagonal target that is off by a factor of 1.3 from the true covariance (dot-dashed), the covariance estimated from 2000 samples (dashed), and the shrinkage estimator that combines the estimated and target covariances to get closer to the correct result (thin solid). The diagonal and off-diagonal parts of the matrix are shrunk separately, giving a shrinkage estimator close to the sample estimate on the diagonal and close to the target on the off-diagonal. We approximate the likelihood as Gaussian in the $C_l$.
}
\label{shrunk}
\end{center}
\end{figure}

The effect of the shrinkage estimator on estimated likelihoods is illustrated in Fig.~\ref{shrunk}, where separate shrinkage estimates are used for the diagonal and off-diagonal parts of the covariance matrix. Here toy samples were generated from Gaussian model with a known true covariance, and the shrinkage target is diagonal and $30\%$ wrong. The shrinkage estimator is significantly closer to the true result than using either the direct sample estimate or the diagonal target, as expected. Using a purely diagonal covariance approximation with the variances estimated from the samples would give a similar result. As more samples are used the shrinkage estimator gradually includes more information about the off-diagonal correlations from the simulations.

\begin{figure}
\begin{center}
\epsfig{figure=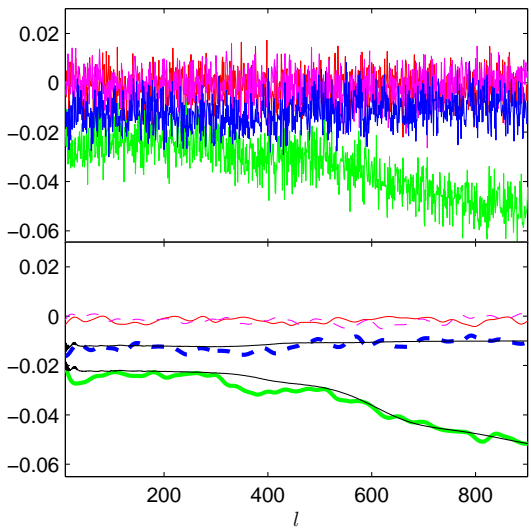,width=7cm}
\caption{Diagonal strips of the maximum-likelihood correlation matrix estimated from 23000 simulations of the WMAP 5-year temperature maps~\cite{Lewis:2008wr}, $(l,l+1)$ [thin solid], $(l,l+2)$ [thick solid], $(l,l+3)$ [thin dashed] and $(l,l+4)$ [thick dashed]; the elements become very small when further away from the diagonal. The top figure is the raw estimate, the bottom figure shows the result smoothed with a Gaussian kernel of width $\sigma_l=10$. The $(l,l+1)$  and $(l,l+3)$ elements are much smaller than $(l,l+2)$, $(l,l+4)$ due to the near north-south symmetry of the foreground mask in galactic coordinates. The thin black solid lines in the bottom half show an analytic approximation for the corresponding parts of the correlation matrix~\cite{Lewis:2008wr}.
}
\label{WMAP_cov}
\end{center}
\end{figure}

The division of the covariance into components with different properties can be further generalized. Cut-sky effects tend mostly to correlate nearby $l$, but the matrix is smooth in $l$, so for example $M_{l(l+1)}$ is similar to  $M_{(l+1)(l+2)}$ (see Fig.~\ref{WMAP_cov}). This suggests slicing the matrix into diagonal strips: $M_{ll}$ (for all $l$),
$M_{l(l+1)}$, $M_{l(l+2)}$, etc, where each strip is given its own shrinkage intensity. In practice $M_{l(l+n)}$ is very small for $n\gg 1$ in most cases (the matrix is nearly band diagonal), so that the combined shrinkage estimator effectively sets $n\gg 1$ elements to the target. Using a shrinkage estimator does however allow the possibility of the estimator including strongly off-diagonal correlations if they appear to be needed by the samples. If the target is nearly zero for $n \gg 1$, then all the $n\gg 1$ elements could be lumped together with a single shrinkage intensity, reducing noise in the intensity due to the small size of the strips in the corners of the matrix. The strips with $n\sim1$ will gradually change from the target to the sample estimator as the number of samples is increased.

\begin{figure}
\begin{center}
\epsfig{figure=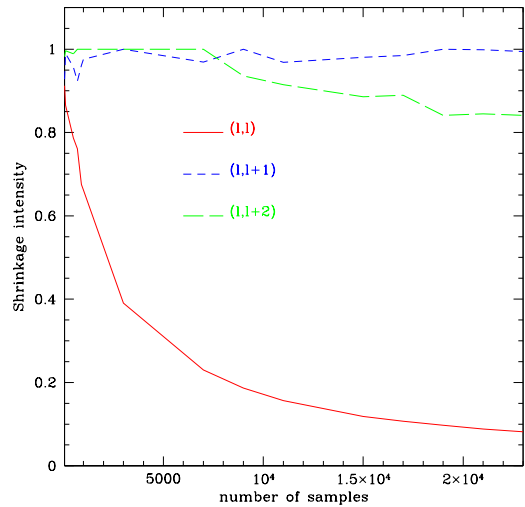,width=8cm}
\epsfig{figure=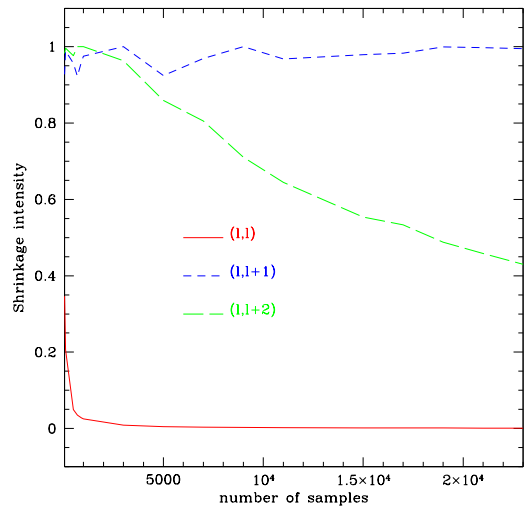,width=8cm}
\caption{The shrinkage intensity $\hat{\lambda}^*$ against the number of WMAP 5-year simulation samples used, using the analytic approximation as the target (top) and using a target that is significantly wrong (bottom; the analytic approximation multiplied by 1.3). The shrinkage intensity is calculated separately for each strip of the normalized matrix, the plots show the diagonal of the covariance (solid), first off-diagonal strip (short-dashed) and the second strip (long-dashed).
}
\label{lambda_scaling}
\end{center}
\end{figure}

Fig.~\ref{WMAP_cov} compares the covariance estimated from 23000 realistic simulations of the WMAP 5-year temperature maps~\cite{Lewis:2008wr} with an analytic approximation for off-diagonal parts of the covariance. The analytic result agrees rather well, indicating that in this case an extremely large number of simulations would be required to improve the off-diagonal result significantly from simulations. On the other hand if there was a disagreement between the simulations and analytic result, a shrinkage estimator would account for this and start to correct the result. On the much larger diagonal analytic approximations are only accurate to $\clo(5\%)$ (see Fig.~\ref{WMAP_diag}), so a relatively modest number of simulations can improve on the analytic result.

Fig.~\ref{lambda_scaling} shows how the shrinkage intensity varies with the number of simulations, both for the case of a significantly wrong target, and the example of WMAP where we have an analytic approximation accurate at the $5\%$-level. The better the target is the more simulations are required to improve on it from the simulations: with a significantly wrong target the simulation estimator rapidly dominates the shrinkage estimator ($\hat{\lambda}^*\rightarrow 0$). With a good analytic result the diagonal estimate is gradually improved over a few thousand simulations, but far more simulations would be needed to correct small inaccuracies in the off-diagonal terms.

A disadvantage of the shrinkage approach is that it does not specify how parameters should be scaled or divided into subsets with different shrinkage intensities. For example it may also be beneficial to divide in ranges of $l$ if the accuracy of the target varies as a function of $l$. This is the case for many power spectrum estimators since the covariance can be calculated accurately in the noise-dominated regime, with the main inaccuracies coming from the sample variance over the acoustic peaks. The extent to which covariance errors affect parameters also varies as a function of $l$: errors in the noise dominated regime have little effect because there is no signal there.

\subsection{Covariance models}

\begin{figure}
\begin{center}
\epsfig{figure=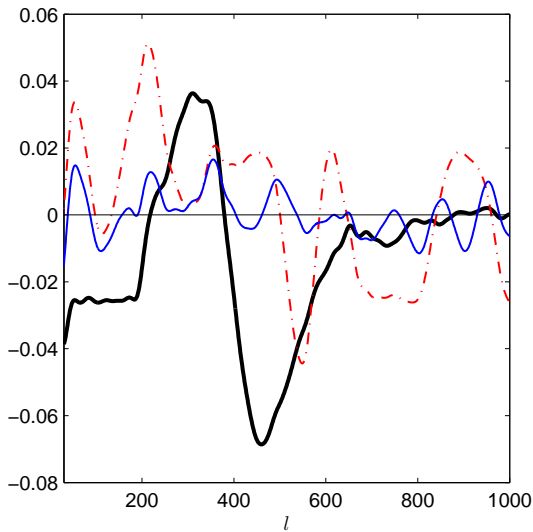,width=7cm}
\caption{The fractional difference of various approximations to the smoothed diagonal of the WMAP5 $C_l$ covariance estimated from 23000 samples from Ref.~\cite{Lewis:2008wr}. The thick solid line shows an analytic approximation, which is accurate at the $5\%$-level. Thin lines show the result from fitting a cubic spline to the $C_l$ variance using 100 (dot-dashed) and 1000 (solid) simulation samples. Spline nodes are separated by $\Delta_l=50$.}
\label{WMAP_diag}
\end{center}
\end{figure}

The general shrinkage estimator uses a target covariance and combines it directly with data from simulations, allowing general deviations from the target to be determined with enough simulations. However in many instances we actually have rather strong priors about the form of the covariance, even without an accurate prior; for example in the case of the CMB the covariance is expected to be strongly diagonally dominated, and also smooth in $l$ unless sharp changes are introduced by the choice of $C_l$-estimator\footnote{Note the WMAP team's analysis does use a sharp cut in $l$ to switch between pseudo-$C_l$ weighting schemes. However each estimator is separately smooth.} (e.g. see Fig.~\ref{WMAP_cov}). If we are confident that this structure is correct, we should be able to use it to improve our covariance estimator.

The approach adopted by the WMAP team is to use an analytic model of the covariance and then calibrate it from simulations by fitting parameters to a model of the covariance diagonal~\cite{Verde:2003ey}. If the model has the right shape this can reduce to fitting a very small number of parameters, which can be done accurately from a relatively small number of simulations, making this a very good approach. On the other hand if the actual behaviour is more complicated than the assumed model, or the off-diagonal correlations cannot be calculated reliably analytically, this could lead to  misestimation of the covariance.


For analysis methods that do not introduce features in $l$, variations in the $C_l$ covariance diagonal are smooth on a scale determined by the scale of the acoustic peaks. Re-scaling the diagonal of the covariance using a smooth model, and taking the correlation matrix from the analytic result, therefore gives close to the right answer; e.g. using the estimator
\begin{equation}
M_{ll'}^{\text{model}} = p_l T_{ll'} p_{l'}
\end{equation}
(no sum), where $\bf{T}$ is the analytic model and $p_l$ are fit to the simulation data using cubic splines. Fig.~\ref{WMAP_diag} shows that fitting only 100 simulation samples using cubic-splines for the diagonal can be as accurate as the analytic approximation. With a thousand or more simulations the covariance diagonal can be determined to the sub-percent level. If required additional parameters could be introduced to model variations in the off-diagonal components. If the theoretical model dependence of the covariance is not well captured by the approximation of Ref.~\cite{Hamimeche:2008ai}, fits could also be made as a function of simulation parameters and then interpolated as required.


An alternative to fitting a model for the covariance when no good model is available would be using a smoothed version of the estimated covariance, for example by applying a Gaussian smoothing kernel to each diagonal strip of the matrix. If the true covariance is indeed smooth, this will significantly reduce the sampling noise and hence improve the covariance estimate. For example applying a smoothing kernel of width $\sigma_l=5$ to the diagonal strips produces results for the likelihood in Fig.~\ref{shrunk} that are similar to the shrinkage estimator, depending on the realization. However it does not guarantee that the smoothed matrix is invertible.

\section{Combining low and high $l$}

\begin{figure}
\begin{center}
\psfig{figure=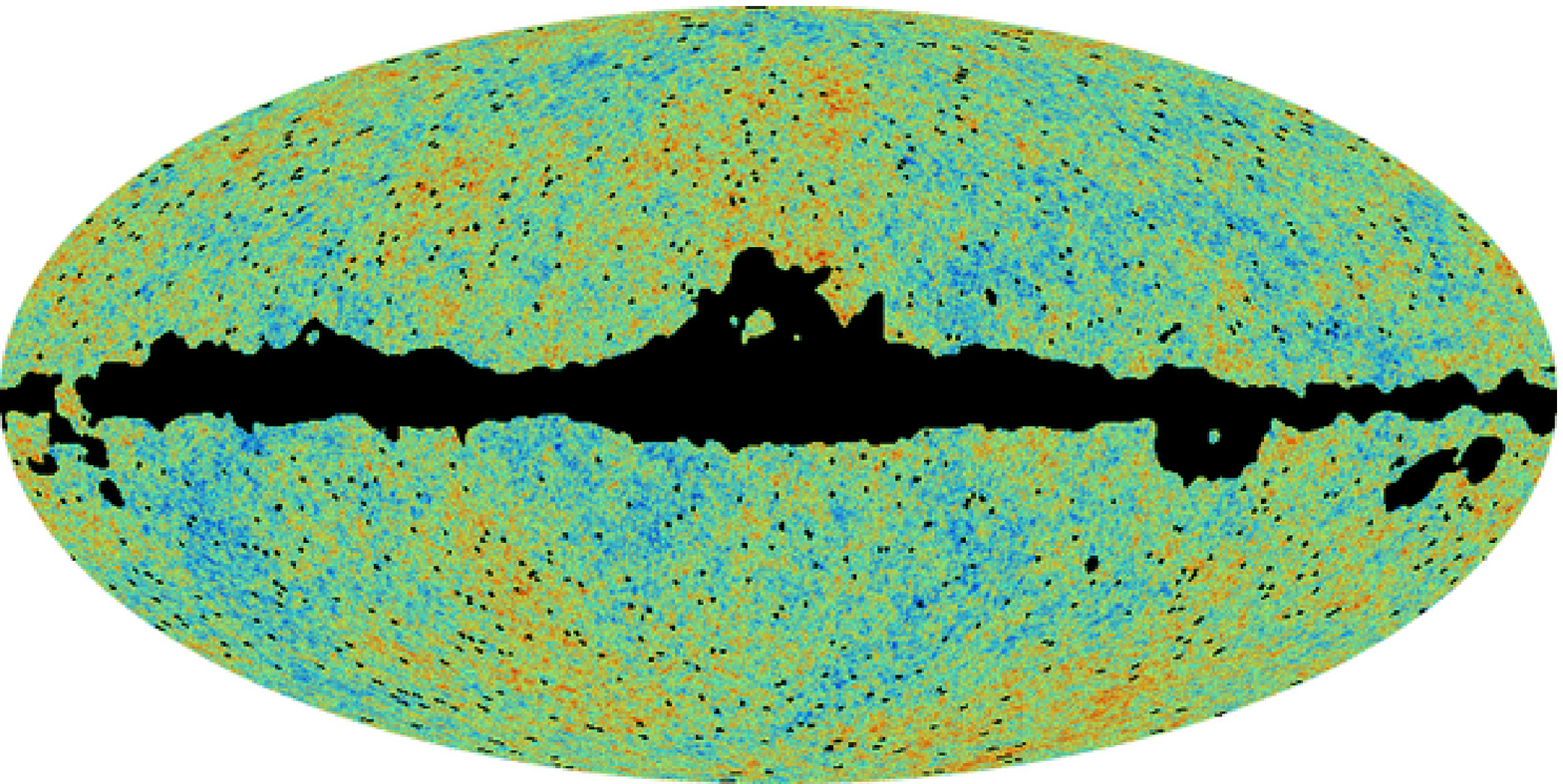,width=9cm}
\psfig{figure=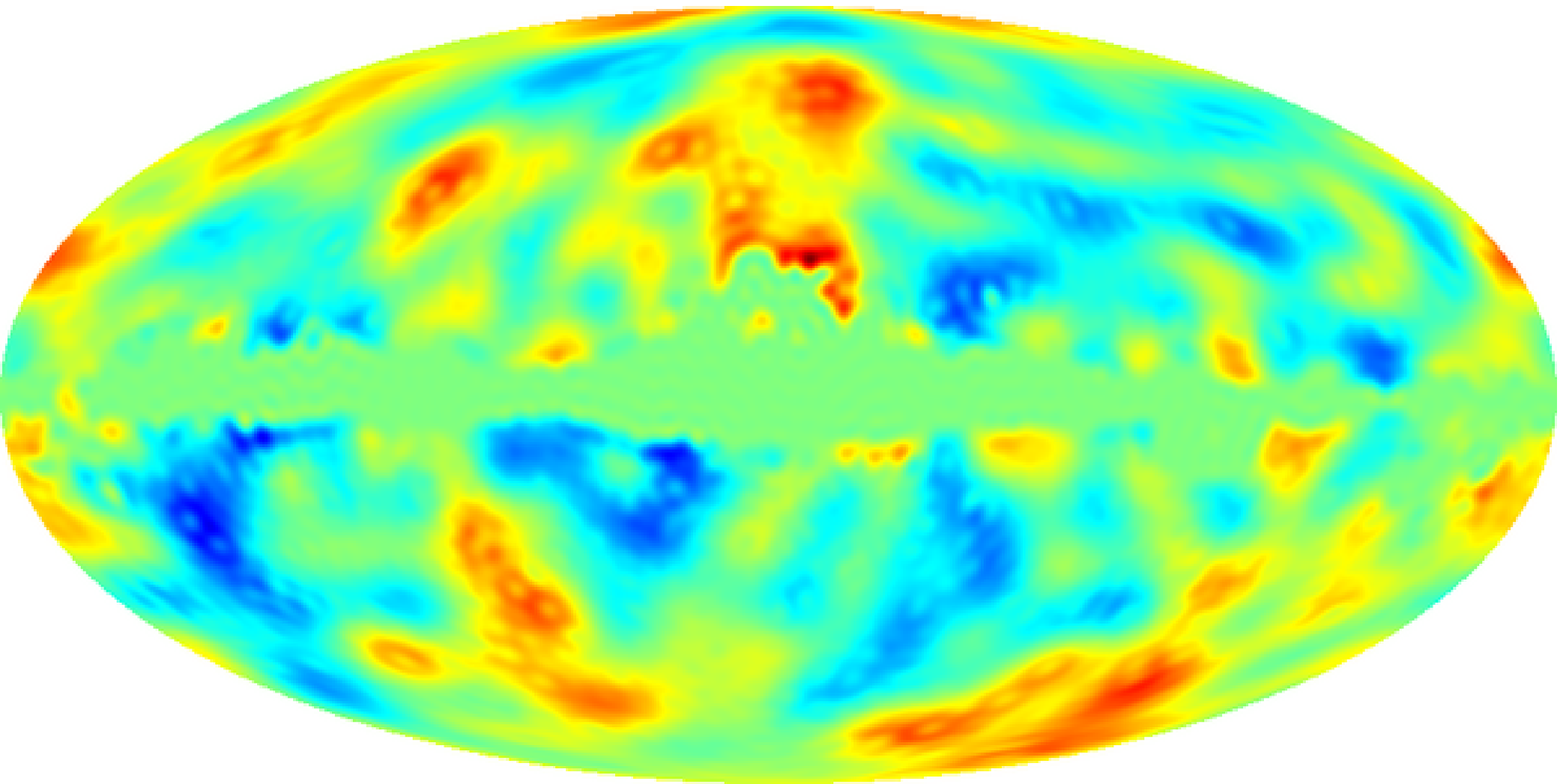,width=9cm}
\psfig{figure=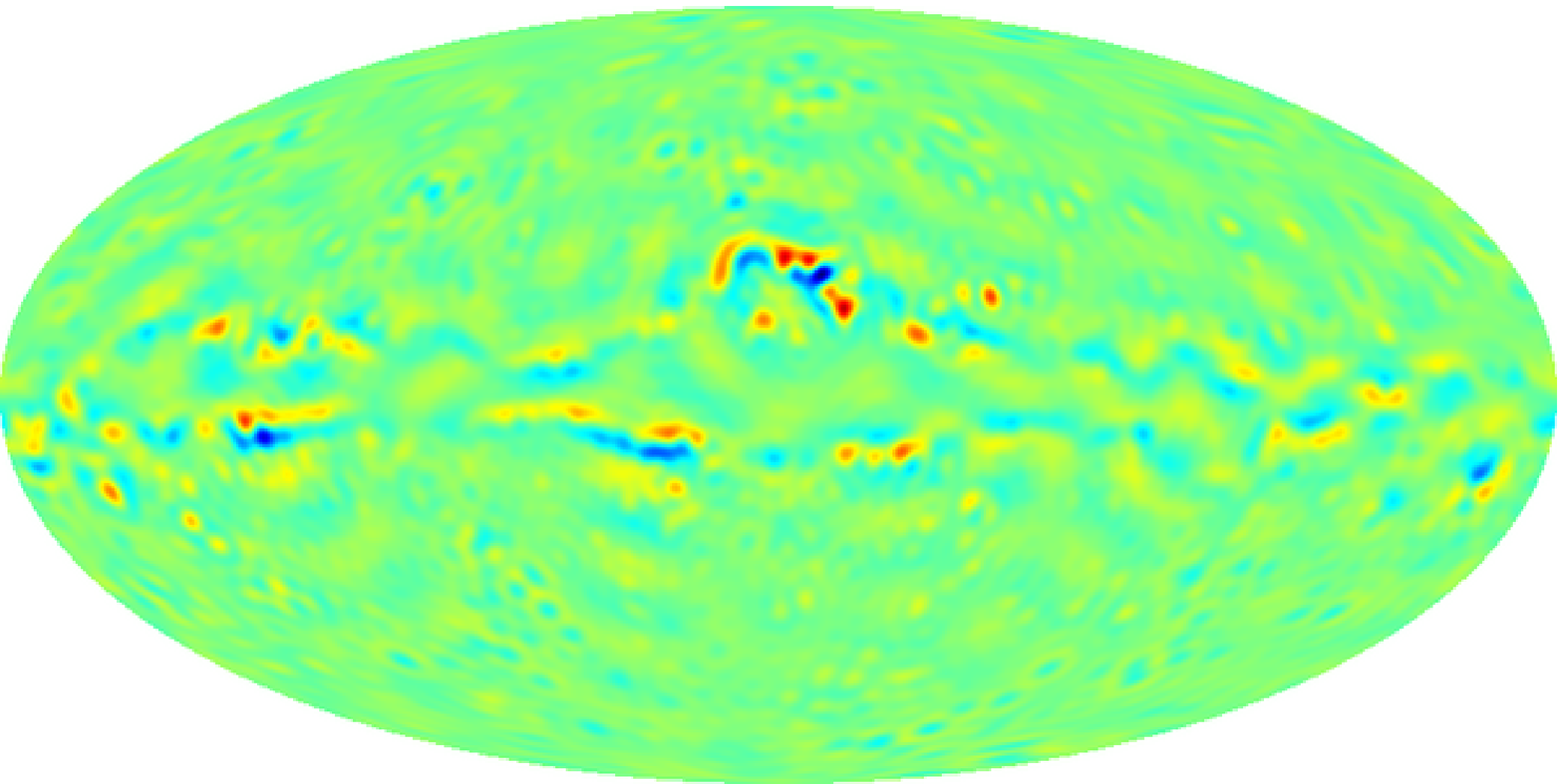,width=9cm}
\caption{Input temperature map with kp2 mask~\cite{Bennett:2003bz}  (top), low-$l$ modes  $\la\tilde{\vX}\tilde{\vX}^\dag\ra\tilde{\mM}^\dag\vX_s$ with $\lexact=20$, $\llow=80$ (middle, see Appendix~\ref{lowl}), and the low-$l$ modes'  contribution from $l>\lexact$ (bottom). Note how the supported modes are going smoothly to zero at the cut edges, and the mixing from to $l>\lexact$ is mostly near the edges of the cut. Colour scales are not the same.
}
\label{maps}
\end{center}
\end{figure}

\begin{figure}
\begin{center}
\psfig{figure=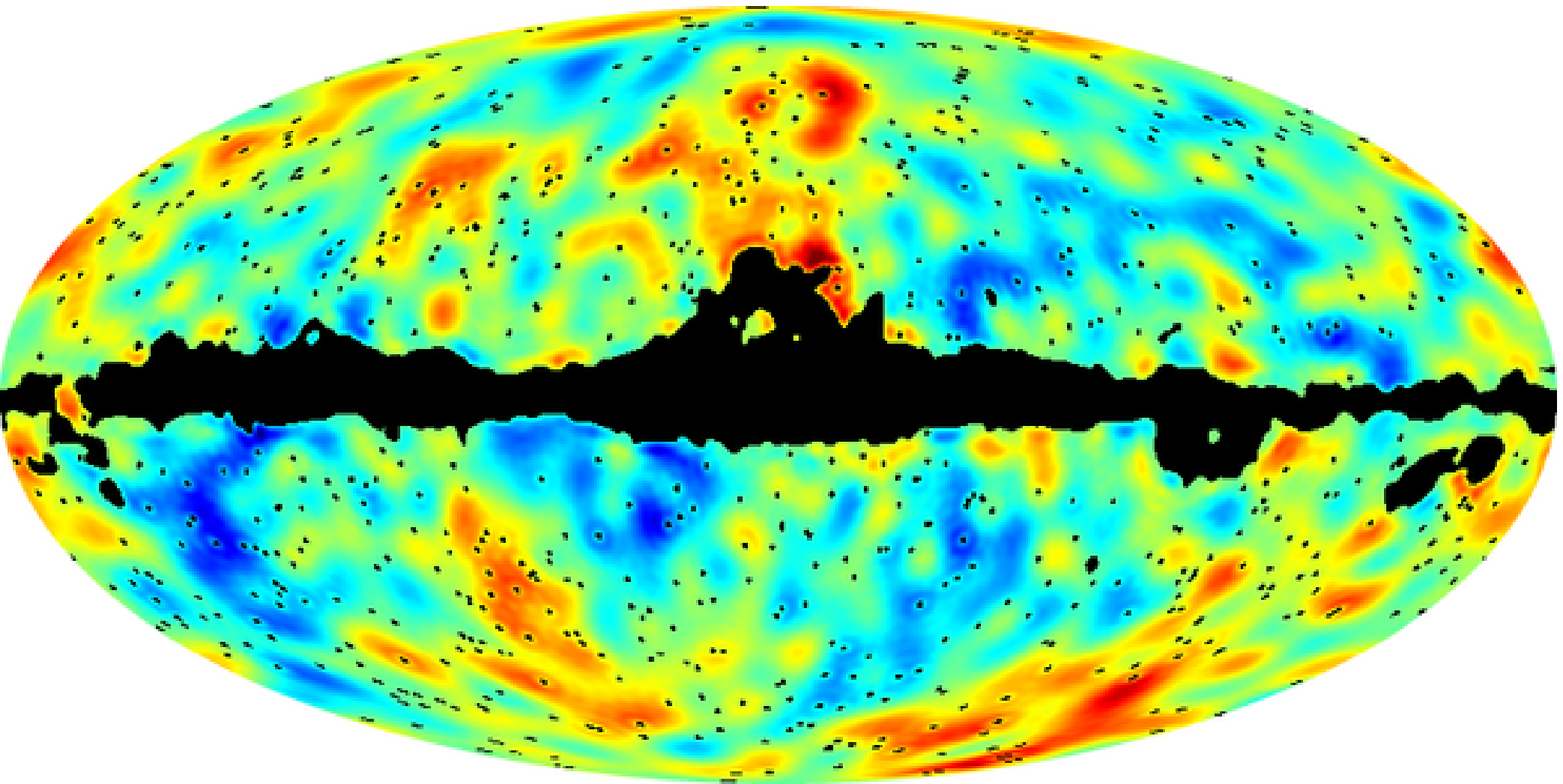,width=9cm}
\psfig{figure=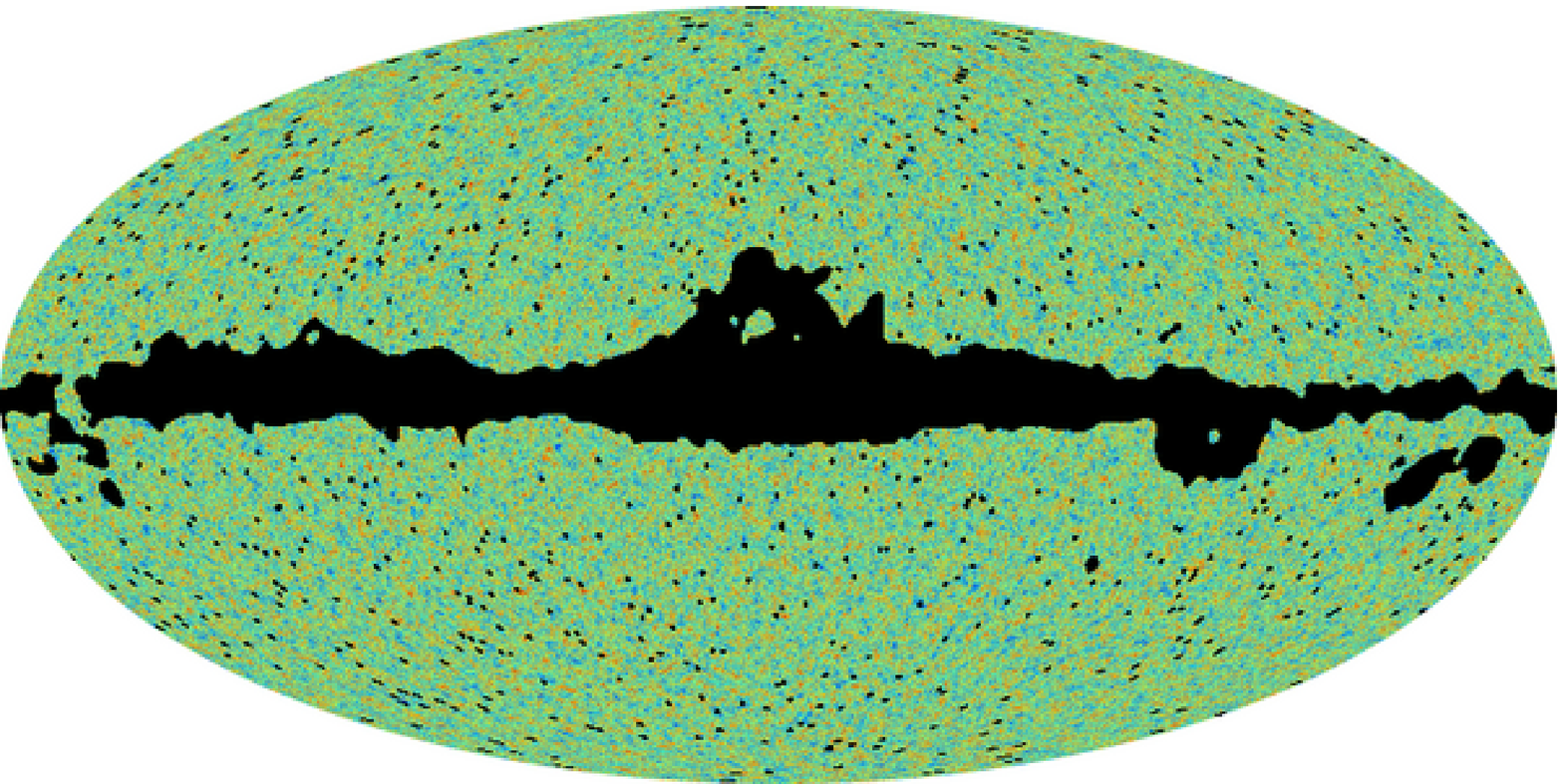,width=9cm}
\psfig{figure=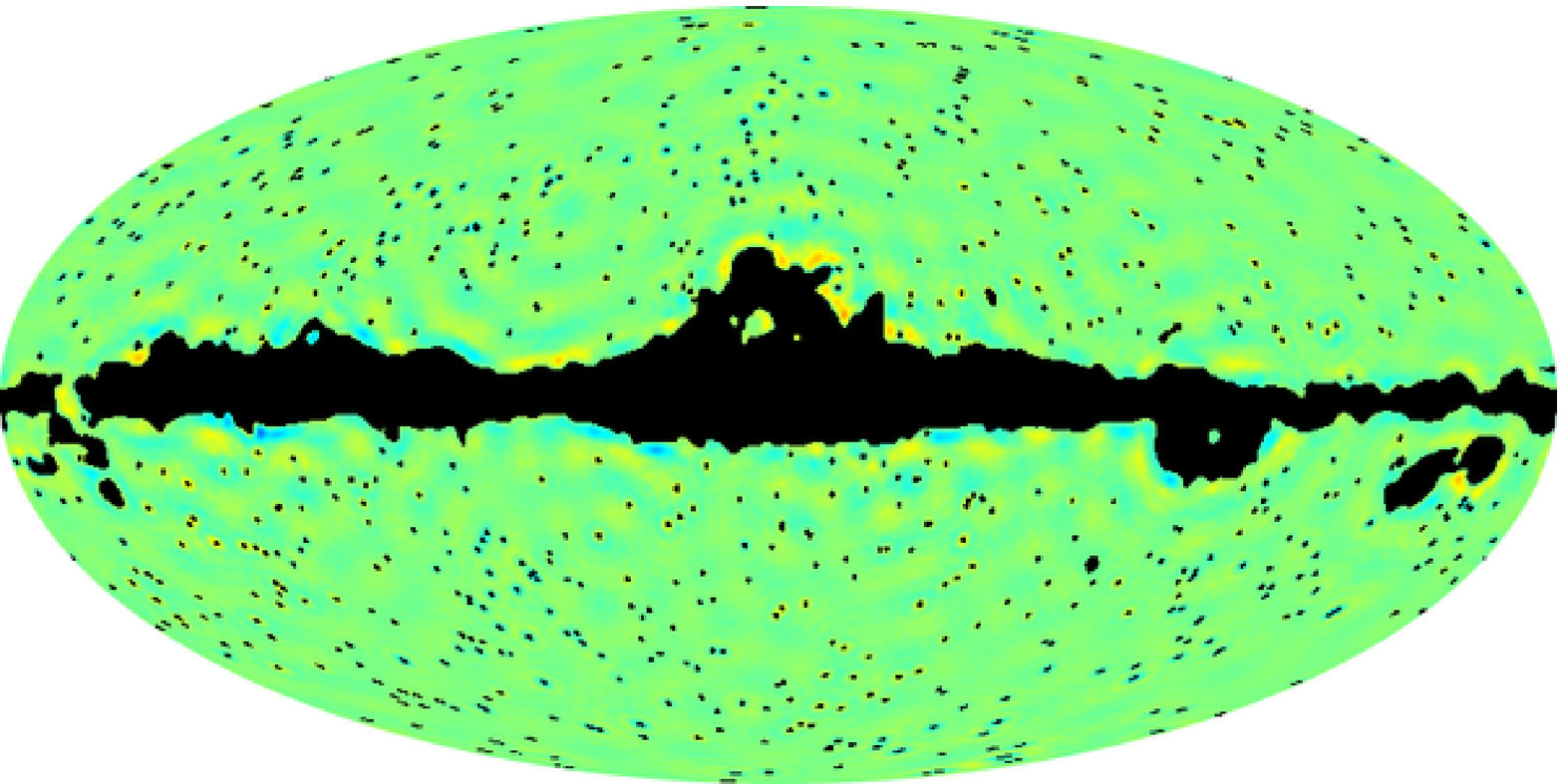,width=9cm}
\caption{Decomposition of the same map into low-$l$ modes (top), independent high-$l$ modes (middle), and the contribution from $l>\lexact$ to the low-$l$ map.  Here $\lexact=30$, $\llow=120$. Note a slightly larger mask would remove much of the high-$l$ contribution.
}
\label{maps2}
\end{center}
\end{figure}
We have discussed how the likelihood can easily be calculated at high $l$ without losing very much information by compressing the data into a set of pseudo-$C_l$ estimators. However at low $l$ one should be able to perform a more optimal analysis, which is desirable since the exact shape of the likelihood function depends on the particular realization of low $l$ modes on the sky. The WMAP data analysis uses a pixel based likelihood at low $l$~\cite{Dunkley:2008ie}, essentially assuming Gaussianity of a set of low resolution pixel values and using the pixel covariance to calculate the likelihood. An alternative would be to work directly in harmonic space, avoiding pixelization issues, as discussed in Appendix~\ref{lowl}. However one calculates the low $l$ likelihood, the question remains of how to combine this with a high-$l$ likelihood, given it will not be possible to do an exact $l$ separation on the cut sky.

Given a set of observed modes on the sky $\tilde{\vX}$, a low $l$ likelihood operates on some subset of the modes given by $\vX_s = \tilde{\mM}\tilde{\vX}$, where $\tilde{\mM}$ is rectangular and $\vX_s$ is sufficiently small that an exact likelihood calculation is numerically feasible. In general $\vX_s$ will contain the low $l$ modes we are aiming to analyse exactly ($l\le \lexact$), along with some leakage from higher $l> \lexact$ due to the sky cut.
It follows that in the fiducial model the remaining information is contained in the independent high-$l$-dominated modes
\begin{eqnarray}
\tilde{\vX}_> &\equiv& \left(\mI -\la\tilde{\vX}\tilde{\vX}^\dag\ra\tilde{\mM}^\dag[\tilde{\mM}\la\tilde{\vX}\tilde{\vX}^\dag\ra\tilde{\mM}^\dag]^{-1}\tilde{\mM}\right)\tilde{\vX} \nonumber
\\&=& \tilde{\vX} - \la\tilde{\vX}\tilde{\vX}^\dag\ra\tilde{\mM}^\dag[\la{\vX}_s{\vX}_s^\dag\ra]^{-1}\vX_s \label{nondiag}\nonumber\\
&\approx&
\tilde{\vX} - \la\tilde{\vX}\tilde{\vX}^\dag\ra\tilde{\mM}^\dag\vX_s,
\end{eqnarray}
where the third line follows for the temperature in the true model if the modes have been decorrelated and normalized to be white. For temperature and polarization the same construction can be used on the large vector $\vX_{TEB}$ in Eq.~\eqref{nondiag}, though there are simplifying advantages to doing temperature and polarization separately. In a different model these modes should also be approximately independent as long as the fiducial model is reasonably accurate.

From the high-$l$ modes $\tilde{X}_>$, one can construct pseudo-$C_l$ (or more optimal) estimators as normal, however the coupling matrix and covariance are more complicated because the weight function is no longer local in pixel space. If the leakage from low to high $l$ is small, these should have roughly the same properties as the normal cut-sky estimators, and hence $C_l$-estimator likelihood approximations can be used. If the leakage is small we could also just ignore the correlation, which is probably fine as long as nothing interesting is happening over the affected $l$ range.

However as shown in Figs.~\ref{maps} and \ref{maps2}, the mixing of modes to higher $l$ dominates around the cut edges, where higher-$l$ modes are needed so that the projected map goes smoothly to zero on the cut. This suggests that by slightly enlarging or apodizing the mask used for the $l\agt \lexact$ pseudo-$C_l$ analysis, most of the modes already included in the low-$l$ likelihood would be removed. This enlarged cut would only be needed for some number of $l\agt \lexact$ where the leakage is significant; at very high $l$  the leakage is negligible so the original cut can be used.
This is a quick and simple solution to separating the scales to a reasonable approximation: use an accurate likelihood at low $l$ (including some modes from higher $l$), use pseudo-$C_l$ estimators with an enlarged mask over intermediate scales where the leakage is important, then at high $l$ use pseudo-$C_l$  estimators with the original mask.

\begin{figure}
\begin{center}
\psfig{figure=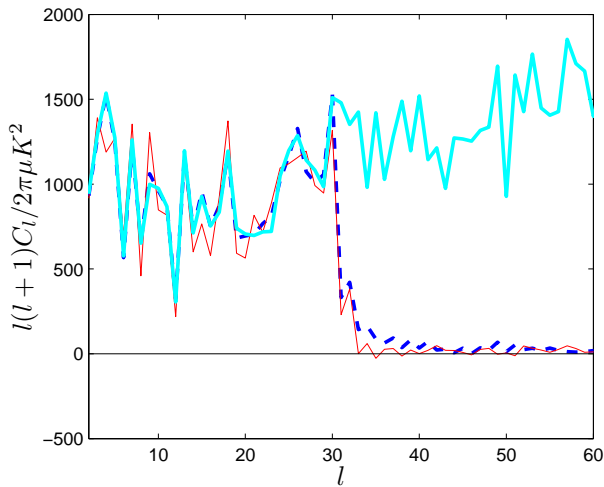,width=8cm}
\caption{
Pseudo-$C_l$ temperature power spectra of a simulated map projected into low-$l$ modes ($\lexact = 30$, $\llow=120$, $\epsilon=0.001$, see the Appendix). Dashed shows the result using the `kp2' mask ($85\%$ of the sky) used to construct the modes. The thin red line shows the estimated $C_l$s using the enlarged `QK75' temperature mask~\cite{Gold:2008kp} that includes only $72\%$ of the sky. The leakage to $l>\lexact$ is relatively small, but is further decreased by using the larger mask. The thick line shows the power spectrum from the cut sky without projecting out the low-$l$ modes.
}
\label{clcuts}
\end{center}
\end{figure}

Fig.~\ref{clcuts} shows that the leakage between scales is small but can further be reduced by using a larger mask over a range $\Delta l \sim 10$ above $\lexact$.
Using an enlarged mask is suboptimal because some modes in the enlarged cut are being lost: the high-$l$ modes included in the low-$l$ likelihood are largely aligned with the boundary, so by cutting all modes in the boundary region transverse modes are lost that could be included. However a small suboptimality is not likely to matter if there is little interesting information in the range $l\agt \lexact$. The separation method has the advantage of keeping the low and high-$l$ analyses straightforward while including almost all of the valuable information at low $l$. The harmonic method described in the appendix is also free from pixelization errors and allows a high-resolution mask to be used: there is no additional loss of information from having to use an enlarged mask and big pixels, as is often the case when using smoothed pixel-based likelihoods. Simply ignoring the correlation between low and high-$l$ would be a good approximation in most models, but is formally incorrect due to double-counting of information in the overlap region.

A recent paper~\cite{Rudjord:2008vc} has claimed that for accurate likelihood calculations on our observed CMB sky, an nearly-exact likelihood is required up to $l\sim 100$. Although this is somewhat surprising given that pseudo-$C_l$ methods are unbiased and not far from optimal, it would still be tractable by direct as well as Gibbs-sampling approaches.


\section{Conclusions}

We have seen that approximate fast power-spectrum likelihoods work well on small scales even with strongly anisotropic noise. The error bars are slightly increased by using a suboptimal method, but since the effect on parameters is small, this is likely to be a price well worth paying for robustness.

The covariance that is required to calculate the high-$l$ likelihood can be calculated from simulations, but in general a very large number of simulations are required to calculate the likelihood accurately. Using approximate analytical results and expected smoothness properties, a more accurate covariance can be calculated with far fewer samples by calibration from simulations. There is a trade off between being able to see any strange behaviour in the simulations and using prior expectations to reduce the number of simulations required. Shrinkage estimators provide one way of combining analytic and simulation results so that if the analytic result is significantly wrong the estimator becomes dominated by the simulation result. Shrinkage estimators however require far more samples than a model-fitting approach,  which can obtain very accurate results from less than a thousand simulations.

With an accurate covariance and set of power spectrum estimators, the high-$l$ likelihood can be calculated quickly and easily. We showed how this could be combined with a more optimal low-$l$ likelihood function, compensating for the small leakage between high and low $l$ while including almost all of the useful information from low $l$. We conclude that a combined likelihood function should be a good option for analysing foreground-cleaned Planck data, though further work is required to model the foreground uncertainties accurately. Fast and robust methods based on power spectra at high $l$ and harmonic near-exact likelihoods at low $l$ are likely to be a very good alternative to globally more accurate (but numerically more difficult) Gibbs sampling methods, as concluded by other authors~\cite{Efstathiou:2003dj,Slosar:2004fr,Hinshaw:2006ia}. In the appendix we described a practical harmonic method for calculating the low-$l$ likelihood without introducing pixelization issues, which can be used easily once the noise covariance is calculated in harmonic space. Allowing for marginalization over foreground templates, or similar method, can often take place in any basis, so generalizing the method suggested here to more realistic cases could be straightforward (see e.g. Ref.~\cite{Efstathiou:2009kt}).

\section{Acknowledgements}
SH gratefully acknowledges the support of the Algerian Ministry of Higher Education and Scientific Research (MESRS) and the British Federation for Women Graduates (FfWG).  AL acknowledges a PPARC/STFC Advanced fellowship and thanks Anthony Challinor, Steven Gratton, and George Efstathiou for discussion. Some of the results in this paper have been derived using the \healpix~\cite{Gorski:2004by} package. We acknowledge the use of the Legacy Archive for Microwave Background Data Analysis (LAMBDA)\footnote{\url{http://lambda.gsfc.nasa.gov/}}. Support for LAMBDA is provided by the NASA Office of Space Science.

\appendix
\section{Covariance priors}
\label{appcov}

The maximum likelihood estimator of the covariance of a set of zero-mean samples of a Gaussian vector $\vn$ is
\begin{equation}
\mCh = \frac{1}{n} \sum_i \vn_i \vn_i^T.
\end{equation}
A $p$-dimensional symmetric positive definite matrix has the Wishart distribution (see e.g. Ref.~\cite{Gupta99}) $\mS \sim W_p(n,\mC)$ if
\begin{equation}
P(\mS) = \frac{|\mS|^{(n-p-1)/2}}{2^{\half n p}\,\Gamma_p(n/2)|\mC|^{n/2}} e^{-\half \Tr(\mS \mC^{-1})},
\end{equation}
so the covariance estimator is distributed as $n\Ch \sim W_p(n,\mC)$. Given the estimator, the true covariance with a flat prior has a distribution $\mC \sim IW_p(n, n\mCh)$ given by
\begin{equation}
P(\mC) = \frac{|\mS|^{(n-p-1)/2}}{2^{\half(n -p-1)p}\,\Gamma_p(\half(n-p-1))|\mC|^{n/2}} e^{-\half \Tr(\mS \mC^{-1})}
\end{equation}
where $n > 2p$.

Assume we wish to calculate the likelihood of $\vd$, where $\vd$ has a $p$-dimensional Normal distribution $\vd \sim N_p(0, \mC)$. Using the estimator of the covariance we want to marginalize out the uncertainty in the true covariance to give
\begin{multline}
L(\vd|\mCh) =
\int \ud \mC \frac{e^{-\half \vd^T \mC^{-1}\vd}}{(2\pi)^{p/2}|\mC|^{1/2}}\times \\
 \frac{|n \mCh|^{(n-p-1)/2}}{2^{\half(n -p-1)p}\,\Gamma_p(\half(n-p-1))|\mC|^{n/2}} e^{-\frac{n}{2}\Tr(\mCh \mC^{-1})}.
\end{multline}
The integral is just another inverse Wishart distribution and can be done, so that
\begin{eqnarray}
-2\log L  &=& (n-p) \log |n\mCh + \vd \vd^T| + \dots \\
&=& (n-p) \log \left( 1 + \frac{1}{n}\vd^T\mCh^{-1}\vd \right) +\dots\\
&\approx& \frac{n-p}{n}\left[ \hat{\chi}^2 - \frac{1}{n} \half (\hat{\chi}^2)^2 + \dots \right] + \dots.
\label{margeapprox}
\end{eqnarray}
(ignoring $\vd$-independent terms) and assuming $\mCh^{-1}$ is invertible ($n\ge p$). Here $\hat{\chi}^2$ ($\sim \clo(p)$ for typical data) is the naive chi-squared one would calculate taking the covariance estimated from simulations to be the true covariance. We are most interested in how the likelihood varies with parameters, so
taking the derivative with respect to a parameter $\theta$ gives
\begin{equation}
\frac{-2\partial \log L}{\partial \theta} \approx \frac{n-p}{n+\hat{\chi}^2} \frac{\partial \hat{\chi}^2}{\partial \theta}.
\end{equation}
Thus using the likelihood $e^{-\hat{\chi}^2/2}$ gives best fit values that agree with the marginalized result, but we need $n\gg p$ samples for the error bars to be similar. If this is not the case then using $\hat{\chi}^2$ will underestimate the error bars by a fraction $\clo([n-p]/[n+p])$; for a fractional accuracy on the error bars better than $\alpha$, we need $n\gg 2 p/\alpha$.
 For the numerical value of the log likelihood to be close we would also need $n\gg p^2$.

One approach to correct for the underestimated error-bars would be to multiply the log-likelihood by an estimate of the under-estimation factor. This should at least ensure the error-bars are consistent, though they would then be larger than could be obtained by using more simulations to estimate the covariance.

If instead of taking a flat prior for $\mC$ we assume an inverse-Wishart prior $IW_p(n_r,\mC_r)$, the result above is the same with
\begin{equation}
\mCh \rightarrow \frac{n \mCh + n_r \mC_r}{n+n_r},
\end{equation}
and $n\rightarrow n+n_r$, which is essentially a shrinkage estimator of the covariance. A hierarchical Bayesian model might take a flat or log prior on $n_r$, or often it is fixed at its minimum value for normalizability.

\section{Cut-sky harmonic low-$l$ likelihood}
\label{lowl}
Here we present a harmonic near-exact low $l$ likelihood method that has no artefacts from using a low-resolution pixelization, and does not rely on a map-smoothing procedure that potentially mixes information inside and outside the foreground mask. The method must work with strongly anisotropic cusped noise as expected with Planck -- but this is no problem in harmonic space. The main disadvantage is that it may be slower than pixel-based codes. For this discussion we neglect all non-ideal complications (foregrounds etc), and assume that if the noise is correlated, the noise covariance can be calculated in harmonic space. The method can be applied to harmonic coefficients computed from high-resolution pixelized maps, or directly to the output of a harmonic map-making algorithm (e.g. Ref.~\cite{ArmitageCaplan:2008re}), thereby avoiding pixelization issues at all stages.

Cut-sky pseudo harmonics $\tilde{\vX}$ are given in terms of the underlying full-sky harmonics $\vX$ and a noise vector $\tilde{\vn}$ by
\be
\tilde{\vX} = \mW^\infty \vX + \tilde{\vn}
\ee
where $\mW^\infty$ is the coupling matrix, which in general is an $n\times\infty$ matrix where we consider $\tilde{\vX}$ only up to the first $n$ modes (ordered in $l$ up to $\llow$).
Elements of the coupling matrix can be
evaluated easily numerically up to moderate $l$
in term of the harmonic coefficients of the window (foreground and point-source mask) $W_{lm}$
using
\begin{multline}
W_{(l_1m_1)(l_2m_2)} =  \\
(-1)^{m_1}\sum_{l} W_{lm}
\sqrt{\frac{(2l_1+1)(2l_2+1)(2l+1)}{4\pi}}\times \quad\quad \\
\threej{l}{l_1}{l_2}{0}{0}{0}
\threej{l}{l_1}{l_2}{m}{-m_1}{m_2}
\end{multline}
where $m = m_1 - m_2$.

We wish to model the covariance using only modes up to $l=\llow$, and aim to compute the exact likelihood for $l\le \lexact$ (where $\lexact \le \llow$). Since $\mW^\infty$ is an $n\times\infty$ matrix, the covariance of $\tilde{\vX}$ depends on all $l$, so we need to project out dependence on $l> \llow$. We define $\mW$ to be the $n\times n$ Hermitian matrix that we can easily compute, so that $\mW^\infty = (\mW, \mY)$ where $\mY$ is $n\times \infty$. For window functions that are $0$ or $1$ everywhere  $\mW^\infty$ is a diagonal projection matrix in pixel space, and in harmonic space completeness of the spherical harmonics then implies $\mW^\infty \mW^\infty{}^\dag  = \mW$ (hence $\mY\mY^\dag = \mW - \mW^2$). Eigenmodes $\ve$ of $\mW$, with $\mW\ve = \lambda\ve$, then have eigenvalues $0\le \lambda\le 1$. Modes with $\lambda \sim 0$ have no signal variance and hence can be deleted without loss of information. The remaining modes that we want have $\ve^\dag \mY \sim 0$, so that dependence on high $l$ is removed.
Since
\be
|\ve^\dag \mY|^2 = \ve^\dag( \mW - \mW^2 ) \ve = \lambda(1-\lambda),
\ee
modes with $\lambda \sim 1$ will have the required properties: we just want the well supported modes
(c.f. Ref.~\cite{Lewis:2003an}). Using sufficiently large $\llow > \lexact$ allows us to construct well supported modes that contain almost all of the information at $l\le \lexact$ (but only incomplete information at $\lexact<l\le\llow$). As $\llow \rightarrow\infty$ we have $\mW^2\rightarrow \mW$, and hence $\lambda \rightarrow\{0,1\}$, so for large enough $\llow$ the supported modes are guaranteed to contain all of the information. Whether or not it works in practice depends on how high $\llow$ needs to be to obtain reliable nearly-optimal results at $l\le \lexact$.

We therefore define the cut-sky supported modes
\be
\vX_c = \hat{\mD}^{-1/2}\hat{\mU}^{\dagger} \tilde{\vX} = \hat{\mD}^{1/2}\hat{\mU}^{\dagger}\vX + \vn_c,
\ee
where $\mW = \mU \mD \mU^\dag$, $\mU$ is orthogonal, and rows of the diagonal matrix $\mD$ corresponding to not well supported modes ($D_{ii}<1-\epsilon_1$) are deleted to form $\hat{\mD}$ and $\hat{\mU}$; $\epsilon_1$ is a free parameter that determines the tolerance for un-modelled mixing of modes with $l>\llow$ and $\vn_c$ is noise. Equalities here and below are taken to hold to order $\epsilon_1$. Note that the uncorrelated noise covariance can be calculated including noise power from all $l$, so $\epsilon_1$ is only determining the signal leakage. For isotropic white noise $\la\vX_c\vX_c^\dag\ra_N = \sigma^2 \mI$, and for low noise levels it may be a good idea to add a small amount of fake isotropic noise $\sigma_\epsilon^2$ so that at $l\agt\llow$ the noise is not negligible and theory leakage becomes small compared to the noise on these scales. Also note that we only need to compute the eigenvectors corresponding to small $\epsilon_1$. In practice $\epsilon_1$ need not be terribly small since for $\llow$ significantly bigger than $\lexact$ the power from $l\ge\lexact$ is suppressed when we focus on modes with interesting signal from $l\le \lexact$.

\begin{figure}
\begin{center}
\psfig{figure=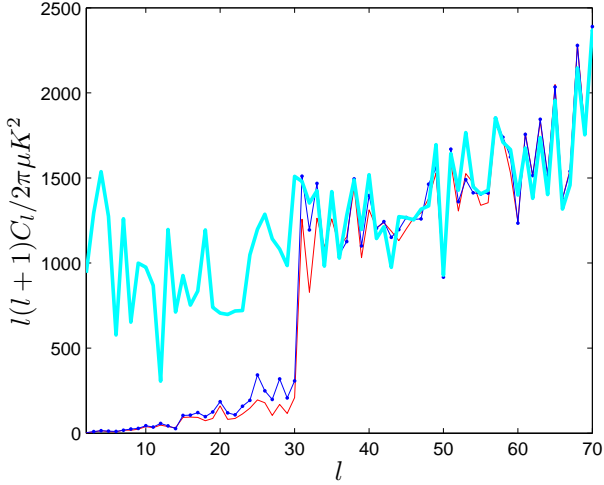,width=8cm}
\caption{
Pseudo-$C_l$ temperature power spectra of a simulated map (`kp2' mask) with the low-$l$ modes ($\lexact=30$) projected out using a signal/noise cut $\epsilon=0.5$ (blue, marked points) and $\epsilon=0.1$ (red, thin line) for $\llow=60$. The thick line is the power spectrum without projection. This shows the trade-off between including all the low-$l$ power and increasing leakage to higher $l$. The leakage between scales is smaller with higher $\llow$ so that more supported modes containing low-$l$ signal can be extracted.
}
\label{projcl}
\end{center}
\end{figure}

Neglecting the small coupling from $l>\llow$, the full covariance of the well-supported modes can be written (and then Cholesky decomposed) as
\be
\la \vX_c \vX_c^\dag\ra = \mS_+ + \mN_S + \mS_{01} + \mN_c +\sigma_\epsilon^2\mI = \mL \mL^\dag.
\ee
Assuming pixel noise is uncorrelated, for the temperature the noise variance is given by $\mN_{lm,l'm'} = \sum_s \Omega_s^2 w(s)^2 \sigma^2(s) Y_{lm}(s)Y_{l'm'}(s)^*$ and $\mN_c = \hat{\mD}^{-1/2}\hat{\mU}^{\dagger} \mN \hat{\mU}\hat{\mD}^{-1/2}$ (the generalization to correlated noise is trivial if $\mN$ can be calculated).
We split $\mS$ into parts, the bit we want from $2\le l\le \lexact$ ($\mS_+$), contributions from other $\lexact<l\le \llow$ ($\mN_S$) that can be though of as part of the noise on the low $l$ modes, and $\mS_{01}$ which is the contribution from any $X$ monopole and dipole (set to be a large number so that the next step projects it out). Then
\be
\mX_L = \mL^{-1}\vX_c
\ee
are uncorrelated with unit variance (in the assumed model), and signal covariance from the target $l$ range is
\be
\la \mX_L \mX_L^\dag\ra_S = \mL^{-1}\mS_+ (\mL^\dag)^{-1} = \mU_s \mD_s \mU_s^\dag.
\ee
Since the $\mX_L$ have variance 1 in total, modes corresponding to small $\mD_s$ will be `noise' dominated. We therefore define the signal to signal plus noise eigenmodes by keeping only modes with $[\mD_s]_{ii} > \epsilon$ for some small $\epsilon$ (for other discussions of similar procedures see Ref.~\cite{Bond:1994aa} and the WMAP likelihood code\footnote{\url{http://lambda.gsfc.nasa.gov/product/map/dr2/likelihood_faster_v2p2p2/wmap_fasttt.ps}}). These new modes contain the interesting signal
\be
\mX_s =  \hat{\mU}_s^\dag \mX_L =   \hat{\mU}_s^\dag \mL^{-1} \vX_c.
\ee
There are typically $\sim (\lexact+1)^2-\lmin^2$ of these per field, corresponding to the number of $X_{lm}$ for $\lmin \le l \le \lexact$. However a significant fraction of the modes have important power from $l\ge \lexact$ due to the sky cut. See Fig.~\ref{projcl} for an example of how the leakage between scales depends on the parameters.

Define $\mM$, $\tilde{\mM}$ as the coupling matrices to the underlying true and pseudo harmonics (at $l\le \llow$), so that
\begin{eqnarray}
\vX_s &=& \mM \vX +\vn_s\equiv  \hat{\mU}_s^\dag \mL^{-1} \hat{\mD}^{1/2}\hat{\mU}^{\dagger} \vX +\vn_s\\
\vX_s &=& \tilde{\mM} \tilde{\vX} \equiv  \hat{\mU}_s^\dag \mL^{-1} \hat{\mD}^{-1/2}\hat{\mU}^{\dagger} \tilde{\vX}.
\end{eqnarray}
Note that $\mM$ should project out any underlying monopole and dipole.
The covariance in the case of the temperature is then given by
\be
\label{final_cov}
\la \vX_s\vX_s{}^\dag\ra = \mN_s + \mM \text{diag}(C_l^{TT}) \mM^\dag
\ee
where
\be
\mN_s = \tilde{\mM} \mN \tilde{\mM}^\dag.
\ee
If desired, and $\mN_s$ is sufficiently non-singular, we can then write $\mN_s = \mL_s \mL_s^\dag$ and define the modes $\mL_s^{-1}\vX_s$ which have unit white noise (in general we can diagonalize even if the matrix is nearly singular). Note that although the signal variance is dominated by modes with $2\le l\le \lexact$, $\mM$ couples in power up to $\llow$.

For the polarization we have
\begin{eqnarray}
\tilde{\vE} &=& \mW_+\vE + i\mW_-\vB \\
\tilde{\vB} &=& \mW_+\vB - i\mW_-\vE,
\end{eqnarray}
where
\begin{multline}
W_{\pm (l_1m_1)(l_2m_2)} =  \\
\half(-1)^{m_1}\sum_{l} W_{lm}
\sqrt{\frac{(2l_1+1)(2l_2+1)(2l+1)}{4\pi}}\times \quad\quad \\
\left[\threej{l}{l_1}{l_2}{0}{2}{-2}\pm
  \threej{l}{l_1}{l_2}{0}{-2}{2}\right]
\threej{l}{l_1}{l_2}{m}{-m_1}{m_2}
\end{multline}
and $m = m_1 - m_2$. Before continuing we change to using real rather than complex $X_{lm}$ modes, as follows.

\subsection{Real harmonics}

It is convenient for numerical work to use real harmonics~\cite{Mortlock00}. For the $T$, $E$, and $B$ we define the real harmonic coefficients
\begin{eqnarray}
X^R_{l|m|} &=& \sqrt{2}\Re( X_{l|m|} ) \qquad X^R_{l-|m|} = \sqrt{2} \Im( X_{l|m|} )
\nonumber\\
X^R_{l0} &=& X_{l0},
\end{eqnarray}
($|m|>0$), which in the full sky case are uncorrelated and have variance $C_l^{XX}$.
The real coupling matrices then relate the pseudo and true harmonics
\begin{eqnarray}
\tilde{\vT}^R &=& \mW^R \vT^R\\
\tilde{\vE}^R &=& \mW_+^R \vE^R + \mW_-^R \vB^R \\
\tilde{\vB}^R &=& \mW_+^R \vB^R - \mW_-^R \vE^R
\end{eqnarray}
where $\mW_+^R$ is symmetric and $\mW_-^R$ is antisymmetric. The real matrices are related to the complex ones by
\begin{eqnarray}
\mW^R_{l|m|l'|m'|} &=& \Re\left( \mW_{l|m|l'|m'|} + (-1)^{m'} \mW_{l|m|l'-|m'|} \right) \nonumber\\
\mW^R_{l-|m|l'|m'|} &=& \Im\left( \mW_{l|m|l'|m'|} + (-1)^{m'} \mW_{l|m|l'-|m'|} \right) \nonumber\\
\mW^R_{l|m|l'-|m'|} &=& \Im\left( -\mW_{l|m|l'|m'|} + (-1)^{m'} \mW_{l|m|l'-|m'|} \right) \nonumber\\
\mW^R_{l-|m|l'-|m'|} &=& \Re\left( \mW_{l|m|l'|m'|} - (-1)^{m'} \mW_{l|m|l'-|m'|} \right) \nonumber\\
\mW^R_{l|m|l'0} &=& \sqrt{2}\Re\left( \mW_{l|m|l'0}\right) \nonumber\\
\mW^R_{l-|m|l'0} &=& \sqrt{2}\Im\left( \mW_{l|m|l'0}\right) \nonumber\\
\mW^R_{l0l'|m'|} &=& \sqrt{2}\Re\left( \mW_{l0l'|m'|}\right) \nonumber\\
\mW^R_{l0l-|m'|} &=& -\sqrt{2}\Im\left( \mW_{l0l'|m'|}\right) \nonumber\\
\mW^R_{l0l'0} &=& \Re\left( \mW_{l0l'0}\right)
\end{eqnarray}
where $|m|, |m'| > 0$ and $\mW$ can be replaced by $\mW_+$ or $i\mW_-$ to obtain the equivalent results for $\mW_\pm^R$. The noise on the real harmonics (for equal and uncorrelated noise on $Q$ and $U$) is given by
\begin{eqnarray}
\la \vE^R (\vE^R)^T\ra_N &=& \la \vB^R (\vB^R)^T\ra_N = \mW_+^{RN} \\
\la \vE^R (\vB^R)^T\ra_N &=& -\la \vB^R (\vE^R)^T\ra_N = \mW_-^{RN}
\end{eqnarray}
where $\mW_-^{RN}$ is evaluated with window function $w_N(s) = \Omega_s w(s)^2 \sigma_P^2(s)$.

\subsection{Polarization modes}

It is convenient to re-complexify the polarization analysis by defining $\vP = \vE^R + i\vB^R$ so that
\be
\tilde{\vP} = (\mW_+^R - i\mW_-^R) \vP \equiv \mW_P \vP
\ee
where $\mW_P$ is Hermitian. The mode construction therefore goes through exactly as for the temperature, except now all matrices are complex, with $\la \tilde{\vP}^*\tilde{\vP}^\dag\ra\ne 0$. The noise covariance under stated assumptions is given by
\be
\la \tilde{\vP} \tilde{\vP}^\dag\ra_N = 2 \mW_P^N \qquad  \la \tilde{\vP}^* \tilde{\vP}^\dag\ra_N=0.
\ee
The real set of modes we end up with is then $\vX_{TEB}\equiv \{\vT^R_s$, $\Re(\vP_s)$, $\Im(\vP_s)\}$, which includes the $E/B$ mixed modes. Nearly-pure $E$ and $B$ would be obtained by keeping only the well supported modes of $\mW_+$ rather than the well supported modes of $\mW_P$ (see Ref.~\cite{Lewis:2003an}). To calculate the signal to noise eigenmodes we can, for example, take $C_l^{EE}=C_l^{BB}$, where $C_l^{EE}$ is a high optical depth model, to ensure that no potentially interesting modes are lost.


\subsection{Implementation}

There is some freedom in how Eq.~\eqref{final_cov} is calculated. $\mN_s$ can be pre-computed assuming we know the noise model and are not fitting it from the data. The coupling matrix $\mM$ has size $\clo(\lexact^2)\times\clo(\llow^2)$. If there is plenty of memory, $\clo(\llow)$ matrices
$\sum_m M_{i(lm)}M_{j(lm)}^*$ can be pre-computed for each $l$, so that calculating the covariance is quick, but taking up $\clo(\lexact^4\llow)$ memory.
Calculating the likelihood is then dominated by the cost of Cholesky decomposition, $\clo(\lexact^6)$, which is quite fast for $\lexact \sim 30$. Alternatively the covariance can be calculated on the fly at a dominating computational cost of $\clo(\lexact^4\llow^2)$ (and only storing $\mM$ of size $\clo(\lexact^2\llow^2)$).

Since information at $l > \lexact$ is subdominant, if $\lexact$ is chosen so that in all models of interest the $C_l$ are of well-determined shape at $l > \lexact$, it may be possible to pre-compute the most time-consuming contribution to the covariance from $\lexact < l \le \llow$, and simply scale it by some weighted average of the spectrum over that $l$-range. Certainly for $\lexact \sim 30$ all the spectra are expected to be very smooth up to $\llow \sim 100$ and this should work well. It could also be fixed at some fiducial model, but there is then a danger of biasing the likelihood from $l\le \lexact$ by misestimating the contribution to the variance from higher $l$.

Using $\lexact = 30$, $\llow=100$, $\epsilon_1 = 0.01$, $\epsilon=10^{-3}$ seems to work well (for Planck a significantly larger $\epsilon_1$ can be used for the temperature since the noise is very small). For an optimal tensor mode analysis we may want to push to $\lexact\sim 150$ to get all the $B$-mode power, which is just about numerically tractable. The low-$l$ harmonic likelihood described here has been implemented in the CosmoMC\footnote{\url{http://cosmologist.info/cosmomc/}} package for parameter estimation since the February 2008 version.

It is possible that the low-$l$ likelihood can be approximated very accurately for very fast subsequent evaluation. For example the likelihood approximation of Ref.~\cite{Hamimeche:2008ai} could be applied to maximum-likelihood power spectrum estimators, or parameters in a likelihood model could be fit to accurately reproduce a full calculation~\cite{Benabed:2009af}. In this case a near-exact low-$l$ method would be an important step for testing or calibrating the approximation, and any speed hit of a harmonic-space approach would be much less important compared to possible accuracy advantages over a pixel-based method. Another possible fast approximation could come from fitting Gibbs sampling results~\cite{Rudjord:2008vc}.

\bibliography{../antony,../cosmomc,samira}

\providecommand{\aj}{Astron. J. }\providecommand{\apj}{Astrophys. J.
  }\providecommand{\apjl}{Astrophys. J.
  }\providecommand{\mnras}{MNRAS}\providecommand{\aap}{Astron. Astrophys.}
\begin{thebibliography}{37}
\expandafter\ifx\csname natexlab\endcsname\relax\def\natexlab#1{#1}\fi
\expandafter\ifx\csname bibnamefont\endcsname\relax
  \def\bibnamefont#1{#1}\fi
\expandafter\ifx\csname bibfnamefont\endcsname\relax
  \def\bibfnamefont#1{#1}\fi
\expandafter\ifx\csname citenamefont\endcsname\relax
  \def\citenamefont#1{#1}\fi
\expandafter\ifx\csname url\endcsname\relax
  \def\url#1{\texttt{#1}}\fi
\expandafter\ifx\csname urlprefix\endcsname\relax\def\urlprefix{URL }\fi
\providecommand{\bibinfo}[2]{#2}
\providecommand{\eprint}[2][]{\url{#2}}

\bibitem[{\citenamefont{Bond et~al.}(2000)\citenamefont{Bond, Jaffe, and
  Knox}}]{Bond:1998qg}
\bibinfo{author}{\bibfnamefont{J.~R.} \bibnamefont{Bond}},
  \bibinfo{author}{\bibfnamefont{A.~H.} \bibnamefont{Jaffe}}, \bibnamefont{and}
  \bibinfo{author}{\bibfnamefont{L.~E.} \bibnamefont{Knox}},
  \bibinfo{journal}{Astrophys. J.} \textbf{\bibinfo{volume}{533}},
  \bibinfo{pages}{19} (\bibinfo{year}{2000}), \eprint{astro-ph/9808264}.

\bibitem[{\citenamefont{Verde et~al.}(2003)}]{Verde:2003ey}
\bibinfo{author}{\bibfnamefont{L.}~\bibnamefont{Verde}} \bibnamefont{et~al.},
  \bibinfo{journal}{Astrophys. J. Suppl.} \textbf{\bibinfo{volume}{148}},
  \bibinfo{pages}{195} (\bibinfo{year}{2003}), \eprint{astro-ph/0302218}.

\bibitem[{\citenamefont{Hamimeche and Lewis}(2008)}]{Hamimeche:2008ai}
\bibinfo{author}{\bibfnamefont{S.}~\bibnamefont{Hamimeche}} \bibnamefont{and}
  \bibinfo{author}{\bibfnamefont{A.}~\bibnamefont{Lewis}},
  \bibinfo{journal}{Phys. Rev.} \textbf{\bibinfo{volume}{D77}},
  \bibinfo{pages}{103013} (\bibinfo{year}{2008}), \eprint{arXiv:0801.0554
  [astro-ph]}.

\bibitem[{\citenamefont{Wandelt et~al.}(2004)\citenamefont{Wandelt, Larson, and
  Lakshminarayanan}}]{Wandelt:2003uk}
\bibinfo{author}{\bibfnamefont{B.~D.} \bibnamefont{Wandelt}},
  \bibinfo{author}{\bibfnamefont{D.~L.} \bibnamefont{Larson}},
  \bibnamefont{and}
  \bibinfo{author}{\bibfnamefont{A.}~\bibnamefont{Lakshminarayanan}},
  \bibinfo{journal}{Phys. Rev.} \textbf{\bibinfo{volume}{D70}},
  \bibinfo{pages}{083511} (\bibinfo{year}{2004}), \eprint{astro-ph/0310080}.

\bibitem[{\citenamefont{Tegmark}(1997)}]{Tegmark:1996qt}
\bibinfo{author}{\bibfnamefont{M.}~\bibnamefont{Tegmark}},
  \bibinfo{journal}{Phys. Rev.} \textbf{\bibinfo{volume}{D55}},
  \bibinfo{pages}{5895} (\bibinfo{year}{1997}), \eprint{astro-ph/9611174}.

\bibitem[{\citenamefont{Wandelt et~al.}(2001)\citenamefont{Wandelt, Hivon, and
  Gorski}}]{Wandelt:2000av}
\bibinfo{author}{\bibfnamefont{B.~D.} \bibnamefont{Wandelt}},
  \bibinfo{author}{\bibfnamefont{E.}~\bibnamefont{Hivon}}, \bibnamefont{and}
  \bibinfo{author}{\bibfnamefont{K.~M.} \bibnamefont{Gorski}},
  \bibinfo{journal}{Phys. Rev.} \textbf{\bibinfo{volume}{D64}},
  \bibinfo{pages}{083003} (\bibinfo{year}{2001}), \eprint{astro-ph/0008111}.

\bibitem[{\citenamefont{Hivon et~al.}(2002)}]{Hivon:2001jp}
\bibinfo{author}{\bibfnamefont{E.}~\bibnamefont{Hivon}} \bibnamefont{et~al.},
  \bibinfo{journal}{\apj} \textbf{\bibinfo{volume}{567}}, \bibinfo{pages}{2}
  (\bibinfo{year}{2002}), \eprint{astro-ph/0105302}.

\bibitem[{\citenamefont{Hansen et~al.}(2002)\citenamefont{Hansen, Gorski, and
  Hivon}}]{Hansen:2002zq}
\bibinfo{author}{\bibfnamefont{F.~K.} \bibnamefont{Hansen}},
  \bibinfo{author}{\bibfnamefont{K.~M.} \bibnamefont{Gorski}},
  \bibnamefont{and} \bibinfo{author}{\bibfnamefont{E.}~\bibnamefont{Hivon}},
  \bibinfo{journal}{Mon. Not. Roy. Astron. Soc.}
  \textbf{\bibinfo{volume}{336}}, \bibinfo{pages}{1304} (\bibinfo{year}{2002}),
  \eprint{astro-ph/0207464}.

\bibitem[{\citenamefont{Efstathiou}(2004)}]{Efstathiou:2003dj}
\bibinfo{author}{\bibfnamefont{G.}~\bibnamefont{Efstathiou}},
  \bibinfo{journal}{Mon. Not. Roy. Astron. Soc.}
  \textbf{\bibinfo{volume}{349}}, \bibinfo{pages}{603} (\bibinfo{year}{2004}),
  \eprint{astro-ph/0307515}.

\bibitem[{\citenamefont{Brown et~al.}(2005)\citenamefont{Brown, Castro, and
  Taylor}}]{Brown:2004jn}
\bibinfo{author}{\bibfnamefont{M.~L.} \bibnamefont{Brown}},
  \bibinfo{author}{\bibfnamefont{P.~G.} \bibnamefont{Castro}},
  \bibnamefont{and} \bibinfo{author}{\bibfnamefont{A.~N.}
  \bibnamefont{Taylor}}, \bibinfo{journal}{Mon. Not. Roy. Astron. Soc.}
  \textbf{\bibinfo{volume}{360}}, \bibinfo{pages}{1262} (\bibinfo{year}{2005}),
  \eprint{astro-ph/0410394}.

\bibitem[{\citenamefont{Hinshaw et~al.}(2007)}]{Hinshaw:2006ia}
\bibinfo{author}{\bibfnamefont{G.}~\bibnamefont{Hinshaw}} \bibnamefont{et~al.}
  (\bibinfo{collaboration}{WMAP}), \bibinfo{journal}{Astrophys. J. Suppl.}
  \textbf{\bibinfo{volume}{170}}, \bibinfo{pages}{288} (\bibinfo{year}{2007}),
  \eprint{astro-ph/0603451}.

\bibitem[{\citenamefont{Efstathiou}(2006)}]{Efstathiou:2006eb}
\bibinfo{author}{\bibfnamefont{G.}~\bibnamefont{Efstathiou}},
  \bibinfo{journal}{Mon. Not. Roy. Astron. Soc.}
  \textbf{\bibinfo{volume}{370}}, \bibinfo{pages}{343} (\bibinfo{year}{2006}),
  \eprint{astro-ph/0601107}.

\bibitem[{\citenamefont{Smith and Zaldarriaga}(2007)}]{Smith:2006vq}
\bibinfo{author}{\bibfnamefont{K.~M.} \bibnamefont{Smith}} \bibnamefont{and}
  \bibinfo{author}{\bibfnamefont{M.}~\bibnamefont{Zaldarriaga}},
  \bibinfo{journal}{Phys. Rev.} \textbf{\bibinfo{volume}{D76}},
  \bibinfo{pages}{043001} (\bibinfo{year}{2007}), \eprint{astro-ph/0610059}.

\bibitem[{\citenamefont{Smith et~al.}(2006)\citenamefont{Smith, Challinor, and
  Rocha}}]{Smith:2005ue}
\bibinfo{author}{\bibfnamefont{S.}~\bibnamefont{Smith}},
  \bibinfo{author}{\bibfnamefont{A.}~\bibnamefont{Challinor}},
  \bibnamefont{and} \bibinfo{author}{\bibfnamefont{G.}~\bibnamefont{Rocha}},
  \bibinfo{journal}{Phys. Rev.} \textbf{\bibinfo{volume}{D73}},
  \bibinfo{pages}{023517} (\bibinfo{year}{2006}), \eprint{astro-ph/0511703}.

\bibitem[{\citenamefont{Oh et~al.}(1999)\citenamefont{Oh, Spergel, and
  Hinshaw}}]{Oh:1998sr}
\bibinfo{author}{\bibfnamefont{S.~P.} \bibnamefont{Oh}},
  \bibinfo{author}{\bibfnamefont{D.~N.} \bibnamefont{Spergel}},
  \bibnamefont{and} \bibinfo{author}{\bibfnamefont{G.}~\bibnamefont{Hinshaw}},
  \bibinfo{journal}{Astrophys. J.} \textbf{\bibinfo{volume}{510}},
  \bibinfo{pages}{551} (\bibinfo{year}{1999}), \eprint{astro-ph/9805339}.

\bibitem[{\citenamefont{{Ashdown} et~al.}(2007)\citenamefont{{Ashdown},
  {Baccigalupi}, {Balbi}, {Bartlett}, {Borrill}, {Cantalupo}, {de Gasperis},
  {G{\'o}rski}, {Heikkil{\"a}}, {Hivon} et~al.}}]{Ashdown:2007ta}
\bibinfo{author}{\bibfnamefont{M.~A.~J.} \bibnamefont{{Ashdown}}},
  \bibinfo{author}{\bibfnamefont{C.}~\bibnamefont{{Baccigalupi}}},
  \bibinfo{author}{\bibfnamefont{A.}~\bibnamefont{{Balbi}}},
  \bibinfo{author}{\bibfnamefont{J.~G.} \bibnamefont{{Bartlett}}},
  \bibinfo{author}{\bibfnamefont{J.}~\bibnamefont{{Borrill}}},
  \bibinfo{author}{\bibfnamefont{C.}~\bibnamefont{{Cantalupo}}},
  \bibinfo{author}{\bibfnamefont{G.}~\bibnamefont{{de Gasperis}}},
  \bibinfo{author}{\bibfnamefont{K.~M.} \bibnamefont{{G{\'o}rski}}},
  \bibinfo{author}{\bibfnamefont{V.}~\bibnamefont{{Heikkil{\"a}}}},
  \bibinfo{author}{\bibfnamefont{E.}~\bibnamefont{{Hivon}}},
  \bibnamefont{et~al.}, \bibinfo{journal}{\aap} \textbf{\bibinfo{volume}{471}},
  \bibinfo{pages}{361} (\bibinfo{year}{2007}), \eprint{astro-ph/0702483}.

\bibitem[{\citenamefont{Lewis and Bridle}(2002)}]{Lewis:2002ah}
\bibinfo{author}{\bibfnamefont{A.}~\bibnamefont{Lewis}} \bibnamefont{and}
  \bibinfo{author}{\bibfnamefont{S.}~\bibnamefont{Bridle}},
  \bibinfo{journal}{Phys. Rev.} \textbf{\bibinfo{volume}{D66}},
  \bibinfo{pages}{103511} (\bibinfo{year}{2002}), \eprint{astro-ph/0205436}.

\bibitem[{\citenamefont{Lewis}(2008)}]{Lewis:2008wr}
\bibinfo{author}{\bibfnamefont{A.}~\bibnamefont{Lewis}},
  \bibinfo{journal}{Phys. Rev.} \textbf{\bibinfo{volume}{D78}},
  \bibinfo{pages}{023002} (\bibinfo{year}{2008}), \eprint{0804.3865}.

\bibitem[{\citenamefont{Reichardt et~al.}(2008)}]{Reichardt:2008ay}
\bibinfo{author}{\bibfnamefont{C.~L.} \bibnamefont{Reichardt}}
  \bibnamefont{et~al.} (\bibinfo{year}{2008}), \eprint{arXiv:0801.1491
  [astro-ph]}.

\bibitem[{\citenamefont{Jones et~al.}(2006)}]{Jones:2005yb}
\bibinfo{author}{\bibfnamefont{W.~C.} \bibnamefont{Jones}}
  \bibnamefont{et~al.}, \bibinfo{journal}{\apj} \textbf{\bibinfo{volume}{647}},
  \bibinfo{pages}{823} (\bibinfo{year}{2006}), \eprint{astro-ph/0507494}.

\bibitem[{\citenamefont{Gupta and Nagar}(1999)}]{Gupta99}
\bibinfo{author}{\bibfnamefont{A.}~\bibnamefont{Gupta}} \bibnamefont{and}
  \bibinfo{author}{\bibfnamefont{D.}~\bibnamefont{Nagar}},
  \emph{\bibinfo{title}{Matrix Variate Distributions}}
  (\bibinfo{publisher}{Chapman \& Hall}, \bibinfo{year}{1999}), ISBN
  \bibinfo{isbn}{1584880465}.

\bibitem[{\citenamefont{{Hartlap} et~al.}(2007)\citenamefont{{Hartlap},
  {Simon}, and {Schneider}}}]{Hartlap:2006kj}
\bibinfo{author}{\bibfnamefont{J.}~\bibnamefont{{Hartlap}}},
  \bibinfo{author}{\bibfnamefont{P.}~\bibnamefont{{Simon}}}, \bibnamefont{and}
  \bibinfo{author}{\bibfnamefont{P.}~\bibnamefont{{Schneider}}},
  \bibinfo{journal}{\aap} \textbf{\bibinfo{volume}{464}}, \bibinfo{pages}{399}
  (\bibinfo{year}{2007}), \eprint{astro-ph/0608064}.

\bibitem[{\citenamefont{Ledoit and Wolf}(2003)}]{Ledoit03}
\bibinfo{author}{\bibfnamefont{O.}~\bibnamefont{Ledoit}} \bibnamefont{and}
  \bibinfo{author}{\bibfnamefont{M.}~\bibnamefont{Wolf}}, \bibinfo{journal}{J.
  Empir. Finance} \textbf{\bibinfo{volume}{10}}, \bibinfo{pages}{603}
  (\bibinfo{year}{2003}),
  \bibinfo{note}{\url{http://www.iew.uzh.ch/chairs/wolf/team/wolf/publications%
/jef.pdf}}.

\bibitem[{\citenamefont{Schaefer and Strimmer}(2005)}]{Schafer05}
\bibinfo{author}{\bibfnamefont{J.}~\bibnamefont{Schaefer}} \bibnamefont{and}
  \bibinfo{author}{\bibfnamefont{K.}~\bibnamefont{Strimmer}},
  \bibinfo{journal}{Statistical Applications in Genetics and Molecular Biology}
  \textbf{\bibinfo{volume}{4}}, \bibinfo{pages}{32} (\bibinfo{year}{2005}),
  \bibinfo{note}{\url{http://www.bepress.com/sagmb/vol4/iss1/art32}}.

\bibitem[{\citenamefont{Pope and Szapudi}(2008)}]{Pope:2007vz}
\bibinfo{author}{\bibfnamefont{A.~C.} \bibnamefont{Pope}} \bibnamefont{and}
  \bibinfo{author}{\bibfnamefont{I.}~\bibnamefont{Szapudi}},
  \bibinfo{journal}{\mnras} \textbf{\bibinfo{volume}{389}},
  \bibinfo{pages}{766} (\bibinfo{year}{2008}), \eprint{0711.2509}.

\bibitem[{\citenamefont{Bennett et~al.}(2003)}]{Bennett:2003bz}
\bibinfo{author}{\bibfnamefont{C.~L.} \bibnamefont{Bennett}}
  \bibnamefont{et~al.} (\bibinfo{collaboration}{WMAP}),
  \bibinfo{journal}{Astrophys. J. Suppl.} \textbf{\bibinfo{volume}{148}},
  \bibinfo{pages}{1} (\bibinfo{year}{2003}), \eprint{astro-ph/0302207}.

\bibitem[{\citenamefont{Dunkley et~al.}(2008)}]{Dunkley:2008ie}
\bibinfo{author}{\bibfnamefont{J.}~\bibnamefont{Dunkley}} \bibnamefont{et~al.}
  (\bibinfo{collaboration}{WMAP}) (\bibinfo{year}{2008}), \eprint{0803.0586}.

\bibitem[{\citenamefont{Gold et~al.}(2009)}]{Gold:2008kp}
\bibinfo{author}{\bibfnamefont{B.}~\bibnamefont{Gold}} \bibnamefont{et~al.}
  (\bibinfo{collaboration}{WMAP}), \bibinfo{journal}{Astrophys. J. Suppl.}
  \textbf{\bibinfo{volume}{180}}, \bibinfo{pages}{265} (\bibinfo{year}{2009}),
  \eprint{0803.0715}.

\bibitem[{\citenamefont{Rudjord et~al.}(2008)}]{Rudjord:2008vc}
\bibinfo{author}{\bibfnamefont{O.}~\bibnamefont{Rudjord}} \bibnamefont{et~al.}
  (\bibinfo{year}{2008}), \eprint{0809.4624}.

\bibitem[{\citenamefont{Slosar et~al.}(2004)\citenamefont{Slosar, Seljak, and
  Makarov}}]{Slosar:2004fr}
\bibinfo{author}{\bibfnamefont{A.}~\bibnamefont{Slosar}},
  \bibinfo{author}{\bibfnamefont{U.}~\bibnamefont{Seljak}}, \bibnamefont{and}
  \bibinfo{author}{\bibfnamefont{A.}~\bibnamefont{Makarov}},
  \bibinfo{journal}{Phys. Rev.} \textbf{\bibinfo{volume}{D69}},
  \bibinfo{pages}{123003} (\bibinfo{year}{2004}), \eprint{astro-ph/0403073}.

\bibitem[{\citenamefont{Efstathiou et~al.}(2009)\citenamefont{Efstathiou,
  Gratton, and Paci}}]{Efstathiou:2009kt}
\bibinfo{author}{\bibfnamefont{G.}~\bibnamefont{Efstathiou}},
  \bibinfo{author}{\bibfnamefont{S.}~\bibnamefont{Gratton}}, \bibnamefont{and}
  \bibinfo{author}{\bibfnamefont{F.}~\bibnamefont{Paci}}
  (\bibinfo{year}{2009}), \eprint{0902.4803}.

\bibitem[{\citenamefont{Gorski et~al.}(2005)}]{Gorski:2004by}
\bibinfo{author}{\bibfnamefont{K.~M.} \bibnamefont{Gorski}}
  \bibnamefont{et~al.}, \bibinfo{journal}{Astrophys. J.}
  \textbf{\bibinfo{volume}{622}}, \bibinfo{pages}{759} (\bibinfo{year}{2005}),
  \eprint{astro-ph/0409513}.

\bibitem[{\citenamefont{Armitage-Caplan and
  Wandelt}(2008)}]{ArmitageCaplan:2008re}
\bibinfo{author}{\bibfnamefont{C.}~\bibnamefont{Armitage-Caplan}}
  \bibnamefont{and} \bibinfo{author}{\bibfnamefont{B.~D.}
  \bibnamefont{Wandelt}} (\bibinfo{year}{2008}), \eprint{0807.4179}.

\bibitem[{\citenamefont{Lewis}(2003)}]{Lewis:2003an}
\bibinfo{author}{\bibfnamefont{A.}~\bibnamefont{Lewis}},
  \bibinfo{journal}{Phys. Rev.} \textbf{\bibinfo{volume}{D68}},
  \bibinfo{pages}{083509} (\bibinfo{year}{2003}), \eprint{astro-ph/0305545}.

\bibitem[{\citenamefont{Bond}(1995)}]{Bond:1994aa}
\bibinfo{author}{\bibfnamefont{J.~R.} \bibnamefont{Bond}},
  \bibinfo{journal}{Phys. Rev. Lett.} \textbf{\bibinfo{volume}{74}},
  \bibinfo{pages}{4369} (\bibinfo{year}{1995}), \eprint{astro-ph/9407044}.

\bibitem[{\citenamefont{Mortlock et~al.}(2002)\citenamefont{Mortlock,
  Challinor, and Hobson}}]{Mortlock00}
\bibinfo{author}{\bibfnamefont{D.~J.} \bibnamefont{Mortlock}},
  \bibinfo{author}{\bibfnamefont{A.~D.} \bibnamefont{Challinor}},
  \bibnamefont{and} \bibinfo{author}{\bibfnamefont{M.~P.}
  \bibnamefont{Hobson}}, \bibinfo{journal}{MNRAS}
  \textbf{\bibinfo{volume}{330}}, \bibinfo{pages}{405} (\bibinfo{year}{2002}),
  \eprint{astro-ph/0008083}.

\bibitem[{\citenamefont{Benabed et~al.}(2009)\citenamefont{Benabed, Cardoso,
  Prunet, and Hivon}}]{Benabed:2009af}
\bibinfo{author}{\bibfnamefont{K.}~\bibnamefont{Benabed}},
  \bibinfo{author}{\bibfnamefont{J.~F.} \bibnamefont{Cardoso}},
  \bibinfo{author}{\bibfnamefont{S.}~\bibnamefont{Prunet}}, \bibnamefont{and}
  \bibinfo{author}{\bibfnamefont{E.}~\bibnamefont{Hivon}}
  (\bibinfo{year}{2009}), \eprint{0901.4537}.

\end{thebibliography}

\end{document}